\documentclass[journal]{IEEEtran}
\usepackage{amsbsy}
\usepackage{floatflt} 

\usepackage{amsmath}
\usepackage{amssymb}
\usepackage{times}
\usepackage{graphics}
\usepackage{graphicx}
\usepackage{xspace}
\usepackage{paralist} 
\usepackage{setspace} 
\usepackage{xypic}
\xyoption{curve}
\usepackage{latexsym}
\usepackage{theorem}
\usepackage{ifthen}
\usepackage{subfigure}
\usepackage{booktabs}
\usepackage{algorithm}
\usepackage{algorithmic}
\usepackage{color}

\newcommand{\diag}{\textup{diag}}

\ifCLASSINFOpdf
\else
\fi

{\theoremheaderfont{\it} \theorembodyfont{\rmfamily}
\newtheorem{theorem}{Theorem}
\newtheorem{lemma}[theorem]{Lemma}

}

\newcounter{brojac}

{\theoremheaderfont{\it} \theorembodyfont{\rmfamily}
\newtheorem{assumption}[brojac]{Assumption}
}

\begin{document}
\title{Linear Convergence Rate of  Class of Distributed Augmented Lagrangian Algorithms}
\author{Du$\check{\mbox{s}}$an Jakoveti\'c, Jos\'e M.~F.~Moura$^{\star}$, and Jo\~ao Xavier
\thanks{The work of the first and third authors was supported by: the Carnegie Mellon$|$Portugal Program under a grant from the Funda\c{c}\~ao de Ci$\hat{\mbox{e}}$ncia e Tecnologia~(FCT) from Portugal; by FCT grants CMU-PT/SIA/0026/2009, FCT PTDC/EMS-CRO/2042/2012, and SFRH/BD/33518/2008 (through the Carnegie Mellon$|$Portugal Program managed by ICTI); by ISR/IST plurianual funding (POSC program, FEDER), and the work of the first and second authors was funded by AFOSR grant~FA95501010291 and by NSF grant~CCF1011903, while the first author was a doctoral or postdoctoral student within the Carnegie Mellon$|$Portugal Program. D. Jakoveti\'c is with University of Novi Sad, BioSense Center, 21000 Novi Sad, Serbia. J.~M.~F.~Moura is with Department of Electrical and Computer Engineering, Carnegie Mellon University, Pittsburgh, PA 15213, USA. J. Xavier is with Instituto de Sistemas e Rob\'otica~(ISR), Instituto Superior T\'ecnico~(IST), University of Lisbon, 1049-001 Lisbon, Portugal. Authors e-mails: djakovet@uns.ac.rs,
moura@ece.cmu.edu, jxavier@isr.ist.utl.pt.}}
%
%
\maketitle
\begin{abstract}
We study distributed optimization where nodes cooperatively minimize the sum  of their individual, locally known, convex costs $f_i(x)$'s,
  $x \in {\mathbb R}^d$ is global. Distributed augmented Lagrangian~(AL) methods have good empirical performance on
  several signal processing and learning applications, but there is limited understanding of their convergence rates and how it depends on the underlying network. This paper establishes globally linear (geometric) convergence rates of a class of deterministic and randomized distributed AL methods,
   when the $f_i$'s are twice continuously differentiable and have a bounded Hessian.
   We give explicit dependence of the convergence rates on the underlying network parameters. Simulations illustrate our analytical findings.

 %
\end{abstract}
\hspace{.43cm}\textbf{Keywords:} Distributed optimization, convergence rate, augmented Lagrangian, consensus.
%
%
%
%
%
%
%
%
\section{Introduction}
\label{section-introduction}
%
%
%
%
%
%
%
%
%
\subsection{Motivation}
\label{subsection-motivation}
We study distributed optimization over a $N$-node, connected, undirected network $\mathcal G = (\mathcal{V}, E)$,
with $\mathcal V$ the set of nodes and $E$ the set of edges.
 Node $i$ has private cost function $f_i(x)$, $f_i:\,{\mathbb R}^d \rightarrow \mathbb R$. We focus on iterative, distributed algorithms that solve the unconstrained problem:
\begin{equation}
\label{eqn-opt-prob-original}
\begin{array}[+]{ll}
\mbox{minimize} & f(x): = \sum_{i=1}^N f_i(x),
\end{array}
\end{equation}
while each node $i$ communicates only with its neighbors. This is the setup in
many applications, e.g., distributed inference, \cite{SoummyaEst},
or distributed source localization, \cite{Rabbat}, in sensor networks.

A popular approach to solve~\eqref{eqn-opt-prob-original},  e.g.,~\cite{Joao-Mota-2,bazerque_lasso,bazerque_sensing,RibeiroADMM1,RibeiroADMM2},
is through the augmented Lagrangian~(AL) dual. The approach assigns a local copy $x_i \in {\mathbb R}^d$ of the global variable
$x$ in~\eqref{eqn-opt-prob-original} to each node $i$, introduces the edge-wise
constraints
$\sqrt{W_{ij}}(x_i-x_j)=0$, $\forall \{i,j\} \in E$,\footnote{We include also self-edges, i.e., $\{i,i\} \in E$, $\forall i$.} and forms an AL dual function by dualizing these constraints
and adding the quadratic penalty $\frac{\rho}{2}\sum_{\{i,j\}\in E,\,i \leq j}W_{ij}\|x_i-x_j\|^2$,
see, e.g., \cite{cooperative-convex}, Section~V, for details\footnote{Here, $\rho \geq 0$ is the penalty parameter
 and $W_{ij}$ are the weights,
collected in the $N \times N$ symmetric matrix $W$,  where
 $W_{ij}>0$ if $\{i,j\} \in E$, $i \neq j$,
$W_{ii}:=1-\sum_{j \neq i} W_{ij}$, and $W$ is  doubly stochastic.}. Denote by $\lambda_{ij} \in {\mathbb R}^d$
 the dual variable that corresponds to the constraint on the edge $\{i,j\}$. Introducing
 the per-node aggregate dual variables $\mu_i:=\sum_{j \in O_i}\sqrt{W_{ij}}\lambda_{ij}\mathrm{sign}(j-i)$,
 where $O_i$ is the node $i$'s neighborhood (including~$i$), one obtains the following dual method to solve~\eqref{eqn-opt-prob-original}:
 \begin{align}
 %
 & \left( \, x_1(k+1),\cdots,x_N(k+1)\,\right) = \nonumber \\
 \label{eqn-update-primal}
 & \mathrm{arg min}_{(x_1,\cdots,x_N) \in {\mathbb R}^{dN}}
 L_a \left( x_1,\cdots,x_N;\,\mu_1(k),\cdots,\mu_N(k)\right) \\
 \label{eqn-update-dual}
 &\mu_i(k+1)  =  \mu_i(k)  +  \alpha  \sum_{j \in O_i} W_{ij} \left( x_i(k+1) - x_j(k+1) \right),
 \end{align}
 where $\alpha>0$ is the (dual) step-size, and $L_a:\,{\mathbb R}^{dN} \times {\mathbb R}^{d N} \rightarrow \mathbb R$,
 is the AL function:
 \begin{align}
 &L_a(x_1,\cdots,x_N;\mu_1,\cdots,\mu_N\,) = \sum_{i=1}^N f_i(x_i) \nonumber \\
\label{eqn-augmented-Lagrangian}
 &+\sum_{i=1}^N \mu_i^\top x_i + \frac{\rho}{2}\sum_{\{i,j\}\in
 E,\,i\leq j}W_{ij}\,\|x_i-x_j\|^2.
 \end{align}
 %
 %
In~\eqref{eqn-update-primal} and~\eqref{eqn-update-dual},
$x_i(k)$ and $\mu_i(k)$ are the node $i$'s primal and dual variables, respectively.
 Dual updates~\eqref{eqn-update-dual} allow for distributed implementation,
 as each node~$i$ needs only the primal variables $x_j(k+1)$ from its immediate neighbors in the network.
 When $\rho=0$, the primal update~\eqref{eqn-update-primal} decouples as well,
 and node $i$ solves for $x_i(k+1)$ locally (without inter-neighbor communications.)
 When $\rho>0$, the quadratic coupling term in~\eqref{eqn-augmented-Lagrangian} (in general) induces
 the need for inter-node communications to iteratively solve~\eqref{eqn-update-primal}.  Many known methods to solve~\eqref{eqn-opt-prob-original} fall into the framework of~\eqref{eqn-update-primal}--\eqref{eqn-update-dual}; see, e.g.,~\cite{Terelius,bazerque_lasso,bazerque_sensing,RibeiroADMM1,RibeiroADMM2,cooperative-convex}.
 These methods are used in various signal processing and learning applications, but, until recently,
 their convergence rates have not been analyzed.
\subsection{Contributions}
\label{subsection-contributions-related-work}
In this paper, we introduce an analytical framework to study the convergence rates
of \emph{distributed} AL methods of type~\eqref{eqn-update-primal}--\eqref{eqn-update-dual} when problems~\eqref{eqn-update-primal} are solved inexactly. While the AL methods that we consider are variations on the existing methods, our analysis
gives new results on the globally linear convergence rates of distributed AL algorithms and brings several important insights
into the performance of distributed multi-agent optimization.

We now explain our technical results. Let $x^\prime(k+1)=(x_1^\prime(k+1)^\top,...,x_N^\prime(k+1)^\top)^\top$ be the solution to~\eqref{eqn-update-primal} when the dual variables
are fixed to~$\mu(k)=(\mu_1(k),...,\mu_N(k))$. Our framework handles arbitrary iterative method that
solves~\eqref{eqn-update-primal}, where the method's initial guess of $x^\prime(k+1)$ (starting point) at iteration~$k$ is set to~$x(k)$. Further, let $\|x(k+1)-x^\prime(k+1)\| \leq {\xi}\, \|x(k)-x^\prime(k+1)\|$, $\forall k$, ${\xi} \in (0,1)$, i.e.,
problem~\eqref{eqn-update-primal} is solved up to a certain accuracy such that
the distance to the solution is reduced~${\xi}$ times with respect to the starting point~$x(k)$.
 Assuming that the cost functions $f_i$'s are twice continuously
differentiable, with bounded Hessian ($h_{\mathrm{min}}I \preceq \nabla^2 f_i(x) \preceq h_{\mathrm{max}} I$, $\forall i$, $\forall x \in {\mathbb R}^d$, $h_{\mathrm{min}}>0$), we give explicit conditions that relate the quantities ${\xi}$, $h_{\mathrm{min}}$, and $h_{\mathrm{max}}$,
and the network's spectral gap $\lambda_2(\mathcal L)$,\footnote{The spectral gap $\lambda_2(\mathcal L)$ is
the second smallest eigenvalue of the weighted Laplacian matrix $\mathcal L:=I-W$.} such that the distributed AL method converges to the solution of~\eqref{eqn-opt-prob-original} at a \emph{globally linear rate}. Furthermore, we explicitly characterize the achieved rate in terms of the above system parameters.

We apply and specialize our results to four iterative distributed AL methods that solve~\eqref{eqn-opt-prob-original} that
mutually differ in how~\eqref{eqn-update-primal} is solved:
deterministic Jacobi, deterministic gradient, randomized Jacobi, and randomized gradient (see Section~{II} for the algorithms' details.)
  We establish with all methods globally linear convergence rates in terms of the total number of per-node communications, and we
  explicitly characterize the rates in terms of the system parameters.
  Furthermore, with deterministic and randomized gradient variants, we establish
  the globally linear convergence rates in terms of the total number of per-node gradient evaluations.

We now highlight several key contributions and implications of our results that distinguish our work from the existing literature on distributed multi-agent optimization.

1. We give a general framework to analyze distributed AL algorithms, and we establish linear convergence rates
for a \emph{wide class} of distributed AL methods. This contrasts with the existing work which typically
studies a specific distributed method, like the distributed ADMM~\cite{ADMMYinJournal,AMMMYinConference}.
In particular, this allows us to establish for the first time linear convergence rates of the distributed AL methods with \emph{randomized} primal variable updates. We remark that, for certain specific methods that we subsume, like the distributed ADMM, the literature gives tighter bounds than we do, as we explain below.

2. To our best knowledge, our results on deterministic and randomized \emph{gradient} variants are the first that establish globally linear convergence rates for any distributed algorithm that solves~\eqref{eqn-opt-prob-original}, \emph{simultaneously} in terms of
  per-node gradient evaluations and per-node communications.

3. We provide distributed methods (deterministic and randomized gradient variants) that involve only simple calculations
(like the gradient-type methods in, e.g.,~\cite{nedic_T-AC}) but achieve significantly faster rates than~\cite{nedic_T-AC}.
That is, we show that through the AL mechanism much faster rates can be obtained compared with respect to
standard distributed gradient methods~\cite{nedic_T-AC}, while maintaining the same communication cost
and similar computational cost per iteration, and requiring additional knowledge on the system parameters.
 Namely,~\cite{WotaoYinDistributedGradient} (see also~\cite{Matei}
 for similar results) studies the method in~\cite{nedic_T-AC} when the costs $f_i$'s are strongly convex and have
 Lipschitz continuous gradients--the setup very similar to ours (We additionally require twice continuously differentiable costs.)
  Assuming that nodes know $h_{\mathrm{max}}$, it shows that
 the distance to the solution after $k$ iterations is $O\left( (1-\alpha\,c_2)^{k/2}+\frac{\alpha h_{\mathrm{max}}}{\lambda_2}\right)$,
 where $\alpha$ is the step-size and $c_2=h_{\mathrm{max}}h_{\mathrm{min}}/(h_{\mathrm{max}}+h_{\mathrm{min}})$.
 From these results, it follows that, to achieve $\epsilon$-accuracy, we need
 $O\left( \frac{\gamma \log(1/\epsilon)}{\epsilon \,\lambda_2}\right)$ per-node communications and
 per-node gradient evaluations, where $\gamma=h_{\mathrm{max}}/h_{\mathrm{min}}$ is the condition number. In contrast, we
 assume with our deterministic gradient
 that nodes know $\lambda_2,h_{\mathrm{min}}$, and $h_{\mathrm{max}}$, and we show that
 the $\epsilon$-accuracy is achieved in $O\left( \frac{\gamma \log(1/\epsilon)}{\,\lambda_2}\right)$
 per-node communications and per-node gradient evaluations.
 In other words, by assuming additional knowledge of $h_{\mathrm{min}}$ and $\lambda_2(\mathcal L)$, we reduce the amount of resources
 needed for $\epsilon$-convergence
  roughly $1/\epsilon$ times, when compared with~\cite{nedic_T-AC}.
\subsection{Related work}
\label{subsection-related-work}
We now further relate our work with the existing literature.
We first consider the literature on distributed multi-agent optimization, and then we consider
the work on the conventional, centralized optimization.

\textbf{Distributed multi-agent optimization}. Many relevant works on this and related subjects have recently appeared.
Reference~\cite{ErminWeiADMM} considers~\eqref{eqn-opt-prob-original} over generic networks as we do,
 under a wide class of generic convex functions.
 The reference shows~$O\left(1/\mathcal K \right)$
  rate of convergence in the number of per-node communications for a distributed ADMM method.
   It is important to note that, differently from our paper,~\cite{ErminWeiADMM} considers generic costs
   for which even in a centralized setting the rates faster than~$O(1/\mathcal K)$
    are not established, while linear rates are not achievable. Reference~\cite{JohanssonDualGradient}
  considers both resource allocation problems and~\eqref{eqn-opt-prob-original} and develops
  accelerated dual gradient methods which are different than our methods. It gives the methods' asymptotic (local) convergence factors
   as~$1-\Omega\left(\sqrt{\frac{\lambda_{\mathrm{min}}(A A^\top)}{\gamma\,\lambda_{\mathrm{max}}(A A^\top)}}\right)$,
   where $A$ is the edge-node incidence matrix and $\lambda_{\mathrm{min}}(\cdot)$
    and $\lambda_{\mathrm{max}}(\cdot)$ denote the minimal non-zero and maximal eigenvalues,
    respectively.\footnote{For two positive sequences $\eta_n$ and $\chi_n$, $\eta_n = \Omega (\chi_n)$ means that $\liminf_{n \rightarrow \infty}\frac{\eta_n}{\chi_n}>0$.} 
    The rates in~\cite{JohanssonDualGradient} are better than the rates that we establish
    for our methods. However,~\cite{JohanssonDualGradient} assumes that each node exactly solves
    certain local optimization problems and is not concerned with establishing
    the rates in terms of the number of gradient evaluations.
    Put differently,~\cite{JohanssonDualGradient} corresponds to exact dual methods (based on the ordinary dual--not AL dual).
    Reference~\cite{Erzeghe-consensus} analyzes
   distributed ADMM for the consensus problem--the special case when $f_i: \,\mathbb R \rightarrow \mathbb R$ is $f_i(x)=(x-a_i)^2$ , $a_i \in \mathbb R$. It establishes the global convergence factor~$1-\Omega(\sqrt{\lambda_2(\mathcal L)})$.
    When we specialize our result to the problem studied in~\cite{Erzeghe-consensus}, their
   convergence factor bound is tighter than ours.
   Finally, references~\cite{ADMMYinJournal,AMMMYinConference}
   analyze the distributed ADMM method therein when the costs are strongly convex
   and have Lipschitz continuous gradients. The method in~\cite{ADMMYinJournal,AMMMYinConference}
   corresponds to our deterministic Jacobi variant when $\tau=1$.
   With respect to our results, the bounds in~\cite{ADMMYinJournal,AMMMYinConference}
   are tighter than ours for the method they study.

References~\cite{Sonia-Martinez,JorgeCortes,ControlApproach,ControlApproach2}
  study distributed primal-dual methods that resemble ours when
 the number of inner iterations $\tau$ is set to one (but their methods are not the same.)
 These works do not analyze the convergence rates of their algorithms.

\textbf{Centralized optimization}. Our work is also related to studies of the AL and related algorithms in conventional, centralized optimization.
There is a vast literature on the subject, and many authors considered
inexact primal minimizations (see~\cite{LanMonteiro,necoaraAL,EcksteinBertsekas} and the references listed in the following paragraphs.)
 Before detailing the existing work, we point to main differences of this paper with respect to usual studies in the literature.
 First, when analyzing inexact AL methods, the literature usually assumes that the primal problems use arbitrary initialization. In contrast,
 we initialize the inner primal algorithm with the previous primal variable. Consequently, our results and the results in the
 literature are different, the algorithms in the literature typically be convergent only to a solution neighborhood, e.g.~\cite{LanMonteiro,necoaraAL}. Second, except for recent papers, e.g.,~\cite{LanMonteiro,necoaraAL}, the analysis of inexact AL is usually done with respect to dual sub-optimality. In contrast, we
 are interested in the primal sub-optimality measures. Third, convergence rates are usually established
 at the outer iteration level, while we--besides the outer iterations level--establish the rates in the number of \emph{inner} iterations.

In summary, we establish \emph{primal sub-optimality} globally linear convergence rates
 in the number of \emph{inner iterations} (overall number of iterations) for our AL methods; such studies
 are not abundant in the literature.

We now detail the literature and divide it into four classes:
 1) ADMM algorithms; 2) AL algorithms; 3) saddle point algorithms; and 4) Jacobi/Gauss-Seidel algorithms. We also point to several interesting connections among different methods.

\textbf{ADMM algorithms}. The ADMM method has been proposed in the 70s~\cite{GabayMerier,GlowinskiMarrocco} and has been since then extensively  studied. References~\cite{TomLuoAL,BertsekasAL,Rockafellar}
show locally linear or superlinear convergence rates of AL methods. Reference~\cite{EcksteinBertsekas}
analyzes convergence of the ADMM method using the theory of maximal set monotone operators, and it
studies its convergence under inexact primal minimizations.
Recently,~\cite{TomLuo,RiceUniv} show that the ADMM method converges globally linearly,
for certain more general convex costs than ours. (The most related work to ours on ADMM is actually
the work on distributed ADMM in~\cite{ADMMYinJournal,AMMMYinConference} that we have already commented on above.)

\textbf{AL algorithms}. Lagrangian duality is classical and a powerful machinery in optimization; see, e.g.~\cite{Urruty} for
general theory, and, e.g.,~\cite{LemarechalUnitCommitment}, for applications in combinatorial optimization and unit-commitment problems.
  The method of multipliers based on the augmented Lagrangian  has been proposed in the late 60s~\cite{Hestenes,ALPowell}.
 The convergence of the algorithm has been extensively studied, also under inexact primal minimizations.
   References~\cite{TomLuoAL,BertsekasAL,Rockafellar}
show \emph{locally} linear or superlinear convergence rates of AL methods.
 The work~\cite{LanMonteiro} analyzes the inexact AL method when
 the primal and dual variables are updated using inexact fast gradient schemes.
 This paper finds the total number of the \emph{inner} iterations needed to achieve an $\epsilon$-accurate primal solution.
 Reference~\cite{necoaraAL} studies AL dual standard and fast gradient methods when the primal problems are solved
 inexactly, up to a certain accuracy~$\epsilon_{\mathrm{in}}$. The reference finds the number of outer iterations and
 the required accuracy~$\epsilon_{\mathrm{in}}$ to obtain an~$\epsilon_{\mathrm{out}}$-suboptimal primal solution.

\textbf{Saddle point algorithms}. This thread of the literature considers iterative algorithms to solve saddle point problems. We divide the saddle point algorithms into two types. The first type of algorithms performs at each iteration only one gradient step with respect to the primal variables.  The second type of algorithms solves at each iteration an optimization problem, like it is done with the AL method in~\eqref{eqn-update-primal}.
We now consider the first type of methods. A classical method dates back to the 50s~\cite{UzawaOld}. In fact,
our distributed gradient AL, when the number of inner iterations is set to $\tau=1$, is an instance of this algorithm.
Reference~\cite{UzawaOld} analyzes stability of the method in continuous time, while~\cite{Golshtein,Maistroskii} analyzes
the method's convergence under diminishing step-sizes. Different versions of the method are considered and analyzed in~\cite{Ruzninski}.
More recently, reference~\cite{NedicSaddlePoint} studies similar algorithms for a wide class of non-differentiable (in general) cost functions and
gives sub-linear rates to a neighborhood of a saddle point (The sub-linear rate is due to the wide function class assumed). In summary,
although one of our algorithms falls into the framework of this class of methods, we could not find the results in the literature that are equivalent to ours.

We now focus on the second type of methods. The classical
method is the Arrow-Hurwitz-Uzawa method in~\cite{UzawaOld}, and since then the algorithm has been thoroughly analyzed and several modifications have been proposed, e.g.,~\cite{UzawaGlobalSuperlinear,SaddlePointInexactLinear,SaddlePointVazno,SaddlePointVaznoInexact,UzawaNovo}.
In fact, our inexact distributed AL method is precisely (an inexact version of) the Arrow-Hurwitz-Uzawa method, applied to a specific saddle point system~(see ahead~\eqref{eqn-saddle-point-system-1}--\eqref{eqn-saddle-point-system-3}.)
This in particular means that the AL algorithm on the dual of~\eqref{eqn-opt-prob-original}, given by~\eqref{eqn-update-primal}--\eqref{eqn-update-dual},
is equivalent to the Arrow-Hurwitz-Uzawa method on a specific saddle point problem~\eqref{eqn-saddle-point-system-1}--\eqref{eqn-saddle-point-system-3}.
 Reference~\cite{SaddlePointVazno} analyzes an exact method therein and establishes its convergence rates.
 References~\cite{SaddlePointInexactLinear,UzawaNovo} analyze
 the inexact methods therein for linear saddle point problems (which corresponds to quadratic cost functions),
 while references~\cite{UzawaGlobalSuperlinear,SaddlePointVaznoInexact}
 analyze inexact methods therein for non-linear saddle point problems (which corresponds to
 more general cost functions.)
 Our analysis is in the spirit closest to this thread of works.
 Although~\eqref{eqn-saddle-point-system-1}--\eqref{eqn-saddle-point-system-3}
  is an instance of the classical setup,
 we could not find in the above literature results equivalent to ours. The main reasons are that
 our inexactness measure is different, and we are interested in counting the number of inner iterations.

\textbf{Jacobi/Gauss-Seidel algorithms}. Our work is also related to studies of
Jacobi/Gauss-Seidel algorithms, in the following sense.
Certain distributed AL methods that we consider solve the \emph{inner problems}~\eqref{eqn-update-primal}
via iterative Gauss-Seidel/Jacobi algorithms. In other words, we employ the
Jacobi/Gauss-Seidel methods at the inner iteration level.
Jacobi and Gauss-Seidel methods have been studied for a long time, e.g.,~\cite{Ortega,Bhaya,BertsekasBook,Asynchronous1,Asynchronous2,Asynchronous3,Asynchronous4,Asynchronous5}.
The methods have been studied both in the synchronous updates setting, e.g.,~\cite{Ortega,BertsekasBook},
and in the asynchronous updates setting, e.g.,~\cite{Bhaya,BertsekasBook,Asynchronous1,Asynchronous2,Asynchronous3,Asynchronous4,Asynchronous5}, in more general setups
than the setup that we consider. Reference~\cite{Ortega} presents, e.g., global convergence
for Jacobi and Gauss-Seidel methods (with cyclic order of variable updates) for solving nonlinear systems $F(x)=0$, $F:\,{\mathbb R}^N \mapsto {\mathbb R}^N$, where $F(x)=A x + \phi(x)$, $A$ is an M-matrix and $\phi$ is a diagonal, isotone mapping (see Theorems
13.1.3. and 13.1.5 in~\cite{Ortega}).
The cyclic Jacobi and Gauss-Seidel methods are known to converge at globally linear rates,
when the gradient of the map $F$ is a diagonally dominant (positive definite) matrix;
see~\cite{BertsekasBook}, Proposition~{2.6.} Reference~\cite{Bhaya} studies
asynchronous multi-node\footnote{Reference~\cite{Bhaya} assumes all-to-all inter-node
communications subject to bounded delays.} iterative methods including Gauss-Seidel and Jacobi, in the presence of
bounded inter-node communication delays. It uses Lyapunov
theory to establish global and local convergence (stability)
of asynchronous iterative methods under various conditions. For example,
it is shown that an asynchronous iterative scheme converges if the local nodes' update maps are block
Lipschitz continuous, and if the corresponding matrix of Lipschitz constants is Schur-stable;
see Theorem~{4.4.4} in~\cite{Bhaya}, other results in Chapter~4, and references therein.
 In contrast with the above existing results, convergence of Jacobi/Gauss-Seidel algorithms in general settings is not our main
 concern; instead, we are interested in the overall AL algorithm where Jacobi/Gauss-Seidel are inner algorithms.
  In contradistinction with the literature, we consider Gauss-Seidel and Jacobi methods for the
special case of minimizing~\eqref{eqn-augmented-Lagrangian}; exploiting this special structure,
we derive \emph{explicit convergence factors}
of the Jacobi/Gauss-Seidel updates. This allows us to explicitly determine the required number of
inner (Jacobi/Gauss-Seidel) iterations~$\tau$ that ensure linear convergence of the overall AL
distributed schemes (See Theorem~1 and Lemmas~5--8 for details).

\textbf{Paper organization}. Section~\ref{section-model-algorithm}
details our network and optimization models and presents
distributed AL methods.
Section~{III} presents our analytical framework
for the analysis of inexact AL and proves the generic result on its convergence rate.
Section~{IV} specializes this result for the four considered distributed methods.
Section~{V}
provides simulations with $l_2$-regularized logistic losses.
  Finally, we conclude in Section~{VI}.

\textbf{Notation}. Denote by: ${\mathbb R}^d$ the $d$-dimensional real space;
 $a_l$ the $l$-th entry of vector $a$;
$A_{lm}$ or $[A]_{lm}$ the $(l,m)$ entry of~$A$; $A^\top$ the transpose of $A$;
 $\otimes$ the Kronecker product of matrices;
$I$, $0$, ${1}$, and $e_i$, respectively, the identity matrix, the zero matrix, the column vector with unit entries, and the $i$-th column of $I$; $J$ the $N \times N$ ideal consensus matrix $J:=(1/N){1} \,{1}^\top$; $\| \cdot \|_l$ the vector (respectively, matrix) $l$-norm of its vector (respectively, matrix) argument; $\|\cdot\|=\|\cdot\|_2$ the Euclidean (respectively, spectral) norm of its vector (respectively, matrix) argument; $\lambda_i(\cdot)$ the $i$-th smallest eigenvalue; $A \succ 0$ means $A$ is positive definite; $\lfloor a \rfloor$ the integer part of a real scalar $a$; $\nabla \phi(x)$ and $\nabla^2 \phi(x)$ the gradient and Hessian at $x$ of a twice differentiable function $\phi: {\mathbb R}^d \rightarrow {\mathbb R}$, $d \geq 1$;
$\mathbb P(\cdot)$ and $\mathbb E[\cdot]$
the probability and expectation, respectively; and
$\mathcal{I}({\mathcal A})$ the indicator of event $\mathcal A$.
For two positive sequences $\eta_n$ and $\chi_n$, $\eta_n = O(\chi_n)$ means that $\limsup_{n \rightarrow \infty}\frac{\eta_n}{\chi_n}<\infty$; $\eta_n = \Omega (\chi_n)$ means that $\liminf_{n \rightarrow \infty}\frac{\eta_n}{\chi_n}>0$; and $\eta_n=\Theta(\chi_n)$ means that $\eta_n = O(\chi_n)$ and $\eta_n = \Omega(\chi_n)$.

\vspace{-4mm}
\section{Distributed Augmented Lagrangian Algorithms}
\label{section-model-algorithm}
The network and optimization models are in Subsection~\ref{subsection-model},
 deterministic distributed AL methods are in Subsection~\ref{subsection-algorithm},
 while randomized methods are in Subsection~{II-C}.
\vspace{-3.5mm}
\subsection{Optimization and network models}
\label{subsection-model}
\textbf{Model}. We consider distributed optimization where $N$ nodes solve the unconstrained
problem~\eqref{eqn-opt-prob-original}.
 The function $f_i: {\mathbb R}^d \rightarrow {\mathbb R}$, known only to node $i$,
has the following structure.
\begin{assumption}[Optimization model]
\label{assumption-bdd-hessian}
The functions $f_i: {\mathbb R}^d \mapsto \mathbb R$
 are convex, twice continuously differentiable with bounded Hessian, i.e.,
 there exist $0<h_{\mathrm{min}} \leq h_{\mathrm{max}}<\infty$, such that, for all $i$:
 \begin{equation}
 \label{eqn-bdd-hessian}
 h_{\mathrm{min}}\,  I \preceq \nabla^2 f_i(x) \preceq h_{\mathrm{max}}\, I,\:\:\forall x \in {\mathbb R}^d.
 \end{equation}
\end{assumption}

Under Assumption~\ref{assumption-bdd-hessian}, problem~\eqref{eqn-opt-prob-original}
 is solvable and has the unique solution $x^\star$. Denote by $f^\star = \inf_{x \in {\mathbb R}^d}f(x) = f(x^\star)$ the optimal value.
  Further, Assumption~\ref{assumption-bdd-hessian} implies
  Lipschitz continuity of the $\nabla f_i$'s and strong convexity of the $f_i$'s, i.e., for all~$i$, $\forall x,y \in {\mathbb R}^d$:
\begin{align*}
& \left\| \nabla f_i(x) -\nabla f_i(y) \right\| \leq h_{\mathrm{max}}\,\|x-y\|,\\
& f_i(y) \geq f_i(x) + \nabla f_i(x)^\top \left( y-x \right) +\frac{h_{\mathrm{min}}}{2}\|x-y\|^2.
\end{align*}

\textbf{Communication model}. We associate with~\eqref{eqn-opt-prob-original} a network $\mathcal{V}$ of~$N$ nodes, described by the graph $\mathcal{G} = (\mathcal{V} ,E),$ where $E \subset \mathcal{V} \times \mathcal{V}$ is the set of edges. (We include self-edges: $\{i,i\}\in E$, $\forall i$.)
\begin{assumption}[Network model]
\label{assumption-network}
The graph $\mathcal{G}$ is connected and undirected.
\end{assumption}

\textbf{Weight matrix and weighted Laplacian}. Assign to graph $\mathcal{G}$
 a symmetric, stochastic (rows sum to one and all the entries are non-negative), $N \times N$ weight matrix $W$,
 with, for $i \neq j$, $W_{ij}>0$ if and only if $\{i,j\} \in E,$
  and $W_{ii}=1-\sum_{j \neq i}W_{ij}$. Let also $\widetilde{W}:=W-J.$
  (See~\eqref{eqn-augmented-Lagrangian} for the role of $W$.)
 We require $W$ to be positive definite and its second largest eigenvalue $\lambda_{N-1}(W)<1.$
 Let $\mathcal L:=I-W$ the weighted graph Laplacian matrix, with $\lambda_2(\mathcal L)=1-\lambda_{N-1}(W) \in [0,1)$ the network spectral gap that
measures how well connected the network is. For example, for
a chain $N$-node network, $\lambda_2(\mathcal L)=\Theta\left(\frac{1}{N^2}\right)$,
while, for expander graphs, it stays bounded away from zero as $N$ grows.

\textbf{Global knowledge assumptions}. We summarize the global knowledge on the system parameters
required by our algorithms beforehand at all nodes.
They all require (a lower bound on) the Hessian lower bound~$h_{\mathrm{min}}$,
(an upper bound on) the Hessian upper bound~$h_{\mathrm{max}}$,
and (a lower bound) on the network spectral gap~$\lambda_2(\mathcal L)$.
In addition, the two randomized methods require (an upper bound) on the number of nodes~$N$.
 Further, each node $i$ initializes its dual variable $\mu_i(0)$ to zero. This is essential
 for the algorithm's convergence. We assume that all nodes initialize their
 primal variables to same values, i.e., $x_i(0)=x_j(0)$, $\forall i,j$; e.g., these are set to zero.
 Equal primal variable initialization is not necessary for convergence but allows for simplified expressions in the analysis.
  In addition, each node knows its neighborhood set~$O_i$ and assigns beforehand the weights
  $W_{ij}$, $j \in O_i$.  We refer to~\cite{arxivVersion} on how all the above global knowledge
  can be acquired in a distributed way. Finally, with all our methods, all nodes use the same algorithm
  parameters: the dual step-size $\alpha$, the AL penalty~$\rho$, the number of inner iterations~$\tau$,
  and the primal step-size~$\beta$ (with gradient algorithm variants). As we will see in Sections~{III} and~{IV}, the
  parameters $\alpha, \beta, \rho$, and $\tau$ need to be appropriately set to ensure convergence; for setting the latter parameters, nodes require knowledge of (bounds on) $h_{\mathrm{min}}$, $h_{\mathrm{max}}$, and $\lambda_2(\mathcal L)$, and also $N$
   with the randomized methods.



\vspace{-3.5mm}
\subsection{Deterministic Methods}
\label{subsection-algorithm}
We present two variants of deterministic distributed AL algorithms of type~\eqref{eqn-update-primal}--\eqref{eqn-update-dual}.
 They differ in step~\eqref{eqn-update-primal}.
Both methods solve~\eqref{eqn-update-primal} through inner iterations,
indexed by $s$, and perform \eqref{eqn-update-dual} in the outer iterations, indexed by $k$.
 With the first variant, nodes update their primal variables
 via a Jacobi method on $L_a(\cdot\,;\mu(k))$ in~\eqref{eqn-augmented-Lagrangian};
 with the second variant, they use a gradient descent method on $L_a(\cdot\,;\mu(k))$.
 At outer iterations~$k$, with both variants,
  nodes update the dual variables via
  the dual gradient ascent method (while the primal variables are fixed).
%
%


\textbf{Jacobi primal updates}. We detail the first algorithm variant.
Later, to present other variants, we indicate only the differences with respect to this one. Denote by: $x_i(k,s)$
   the node $i$'s primal variable at the inner iteration $s$
    and outer iteration $k$; and $\mu_i(k)$ the node $i$'s dual variable
     at the outer iteration $k$.
    Further, as in~\eqref{eqn-update-primal}--\eqref{eqn-update-dual}, denote
    by $x_i(k+1)$
     the node $i$'s primal variable at the end of
     the $k$-th outer iteration.
     We relate the primal variables
     at the inner and outer iterations: $x_i(k,s=0):=x_i(k)$, and $x_i(k+1):=x_i(k,s=\tau)$.
     In addition, nodes
  maintain a weighted average of their own and the neighbors'
   primal variables $\overline{x}_i(k,s):=\sum_{j \in O_i}W_{ij}\,x_j(k,s).$
   Recall that $O_i = \{j\in \{1,\cdots,N\}:\,W_{ij}>0\}$
    is the neighborhood set of node $i$, including node $i$.

The algorithm has, as tuning parameters, the weight matrix $W$,
the number of inner iterations per outer iteration $\tau$,
   the AL penalty parameter $\rho \geq 0$,
   and the dual step-size $\alpha>0$.
The algorithm is in Algorithm~1.
     {\small{
\begin{algorithm}
\caption{AL with Jacobi updates}
\begin{algorithmic}[1]
{\small{
    \STATE (\textbf{Initialization}) Node $i$ sets $k=0$, $x_i(k=0) \in {\mathbb R}^d$, $\overline{x}_i(k=0)=x_i(0)$, and $\mu_i(k=0)=0$.
        \STATE (\textbf{Inner iterations}) Node cooperatively run the Jacobi method for
        $s=0,1,\cdots,\tau-1$, with ${x}_i(k,s=0):=x_i(k)$ and $\overline{x}_i(k,s=0):=\overline{x}_i(k)$:
        {\small{
         \begin{align}
         &x_i(k,s+1) =
         \mathrm{argmin}_{x_i \in {\mathbb R}^d} (f_i(x_i) \nonumber \\
         \label{eqn-AL-algorithm-1}
          &+ \left( \mu_i(k) - \rho \overline{x}_i(k,s) \right)^\top x_i  + \frac{\rho \|x_i\|^2}{2})\\
          \label{eqn-AL-algorithm-2}
         \hspace{-.3cm}
         \overline{x}_i(k,s+1)& = \sum_{j \in {O_i}} W_{ij}x_j(k,s+1),
          \end{align}}}
          and set $x_i(k+1)\!:=\!{x}_i(k,s\!=\!\tau)$, $\overline{x}_i(k+1)\!=\!\overline{x}_i(k,s\!=\!\tau)$.
        \STATE (\textbf{Outer iteration}) Node $i$ updates the dual variable $\mu_i(k)$:
        {\small{
        \begin{align}
             \label{eqn-AL-algorithm-3}
             \mu_i(k+1) &=  \mu_i(k) + \alpha \,\left(x_i(k+1) - \overline{x}_i(k+1) \right).
        \end{align}}}
        \STATE Set $k \mapsto k+1$ and go to step 2.}}
\end{algorithmic}
\vspace{-0mm}
\end{algorithm}}}
Algorithm~1 has outer iterations $k$ (step~3) and inner iterations~$s$ (step~2).
 At inner iteration $s$, $s=0,\cdots,\tau-1$,
  node~$i$ solves the local
  optimization problem~\eqref{eqn-AL-algorithm-1} to obtain $x_i(k,s+1)$,
  broadcasts $x_i(k,s+1)$ to all its neighbors $j \in {O_i}-\{i\}$,
  receives $x_j(k,s+1)$, for all $j \in {O_i}-\{i\}$; and computes $\overline{x}_i(k,s+1)$
   via~\eqref{eqn-AL-algorithm-2}.
   At outer iteration $k$,
   node $i$ updates
     $\mu_i(k)$ via~\eqref{eqn-AL-algorithm-3}. (Note that~\eqref{eqn-AL-algorithm-3}
     is equivalent to~\eqref{eqn-update-dual}.)
     Each inner iteration requires
      one ($d$-dimensional) broadcast transmission per node,
      while the outer (dual) iterations do not require communication.
   Overall, node $i$ performs $\tau$
   broadcast transmissions per~$k$.


\textbf{Gradient-type primal updates}.
This algorithm variant is very similar to the Jacobi variant. It replaces
in the Jacobi variant, Algorithm~1,
 the Jacobi update~\eqref{eqn-AL-algorithm-1} with
the gradient descent update on $L_a(\dot;\,\mu(k))$ in~\eqref{eqn-augmented-Lagrangian}.
After algebraic manipulations, obtain the update:
\begin{align}
& x_i(k,s+1) = \left( 1-\beta\,\rho \right)\,x_i(k,s) + \beta\,\rho \,\overline{x}_i(k,s) \nonumber \\
\label{eqn-update-gradient-descent}
 &-\beta\,\left( \,\mu_i(k) + \nabla f_i(x_i(k,s)) \,\right),
\end{align}
where $\beta>0$ is the (primal) step-size parameter.
Hence, in addition to $W$, $\alpha$, and $\rho$,
the gradient primal update algorithm has an additional tuning parameter~$\beta$.

\subsection{Randomized Methods}
\label{section-randomized-AL}
%
%
%
%
%
%
 We introduce two variants of the randomized distributed AL methods of type~\eqref{eqn-update-primal}--\eqref{eqn-update-dual}.
 Both utilize the same communication protocol, but
 they differ in the way primal variables are updated.
Like the deterministic counterparts, they both update the dual variables
at the outer iterations $k$, and they update the primal variables
 at the inner iterations $s$. At each inner iteration $s$,
 one node, say $i$, is selected
 uniformly at random from the set of nodes $\{1,2,\cdots,N\}$.
 Upon selection, node $i$ updates its primal variable and broadcasts it to all
 its neighbors. We now detail the time and communication models.
  The outer iterations occur at discrete
  time steps of the physical time;
  $k$-th outer iteration occurs at time $\tau\,k$, $k=1,2,\cdots$,
  i.e., every $\tau$ time units.
 We assume that all nodes have synchronized clocks for the dual variable updates (dual variable clocks). Each node $i$
  has another clock (primal variable clock) that ticks according to
  a Poisson process with rate~$1$; on average,
  there is one tick of node $i$ in the time
  interval of width $1$. Whenever node $i$'s Poisson
  clock ticks, node~$i$ updates its primal variable and broadcasts it to neighbors.
  The Poisson process clocks are independent.
  Consider the Poisson process clock that ticks
  whenever one of the nodes' clocks ticks. This process
  is a rate-$N$ Poisson process. Hence, in the time interval of length $\tau$, there are on average
   $\tau\,N$ ticks (primal updates), out of which $\tau$ on average
    are done by $i$. One primal update here corresponds to an update of a \emph{single} node.
     Thus, roughly, $N$ updates (ticks) here correspond to one update (inner)
      iteration of the deterministic algorithm.

More formally, let $(\Theta,\mathcal F, \mathbb P)$ be a probability space.
Let $\{\mathcal{T}_i(a,b]\}_{0\leq a \leq b < \infty}$ be a Poisson process with rate $1$, $i=1,\cdots,N$. (This is the
node $i$'s clock for primal variables.) Thus, for a fixed $a,b$, $\mathcal{T}_i(a,b]:\,\Theta \rightarrow \mathbb R$,
  $\mathcal{T}_i(a,b] = \mathcal{T}_i((a,b]\,;\omega)$, $\omega \in \Theta$, is a Poisson random variable with mean $(b-a)$.
Assume the processes $\mathcal{T}_i$ are independent. Let
$\mathcal{T}$ be a Poisson process defined by $\mathcal{T}(a,b]:=\sum_{i=1}^N \mathcal{T}_i(a,b]$.
   Define the random variable
 $\tau(k):=\mathcal{T}(k\tau, (k+1)\tau]$ (the number of
 ticks across all nodes in the $k$-the outer iteration.)
  Consider the events $\mathcal{A}_{k,j}:=\left\{ \omega \in \Theta:\,
\tau(k;\omega) = j\right\}$, $j=0,1,2,\cdots$.
For $j \geq 1$, define the maps: ${\hat{\imath}}(k,s):\,\mathcal{A}_{k,j} \rightarrow \{1,2,\cdots,N\}$,
$s=0,\cdots,j-1$, by ${\hat{\imath}}(k,s;\omega) = i$, if the $(s+1)$-th tick of $\mathcal{T}$
in the interval $(k\tau,(k+1)\tau]$ comes from node $i$'s clock $\mathcal{T}_i$.

We present two variants of the randomized
  distributed AL algorithm: one updates the primal variables via a Gauss-Seidel method and the other
replaces the Gauss-Seidel updates by gradient updates.

\textbf{Gauss-Seidel updates}.
 The dual variables are updated (instantaneously) at times $k \tau$, $k=0,1,\cdots$.
 We denote by $x_i(k):=x_i(k\tau)$ the node $i$'s
 primal variable at time $k\tau$, $k=0,1,\cdots$
 Further, consider $\omega \in \mathcal{A}_{k,j}$: the total number of ticks $\tau(k)$ of $\mathcal{T}$
  in the interval $(k\tau,(k+1)\tau]$ equals $j$,
  and hence we have $j$ inner iterations (ticks) at the outer iteration $k$.
  For any $\omega \in \mathcal{A}_{k,j}$,
  we denote by $x_i(k,s)$ the node $i$'s variable
  after the $s$-th inner iteration, $s=1,\cdots,j$, $j \geq 1$.
   Also, denote by $x_i(k,0):=x_i(k)$, and, for $\omega \in \mathcal{A}_{k,j}$, $x_i(k,\tau(k)=j):=x_i(k+1)$.
  Each node maintains: 1) the primal variable $x_i(k)$; 2) the dual variable $\mu_i(k):=\mu_i(k\tau)$;
  3)~the (weighted) sum of the neighbors' variables $\overline{x}_i(k):=\sum_{j \in O_i}W_{ij}x_j(k)$; and
 4)~the analogous intermediate variables $x_i(k,s)$ and $\overline{x}_i(k,s)$
   during the inner iterations~$s$. The algorithm is Algorithm~3.
  %
  %
  \begin{algorithm}
\caption{Randomized distributed AL with Gauss-Seidel updates}
\begin{algorithmic}[1]
{\small{
    \STATE (\textbf{Initialization}) Node $i$ sets $k=0$, $x_i(k=0) \in {\mathbb R}^d$, $\overline{x}_i(k=0)=x_i(k=0)$, and $\mu_i(k=0)=0$.
        \STATE (\textbf{Inner iterations}) Set ${x}_i(k,s=0):=x_i(k)$, $\overline{x}_i(k,s=0):=\overline{x}_i(k)$, and $s=0$.
        If $\omega \in \Theta$ is such that $\tau(k)=\tau(k;\omega)>0$, then, for $s=0,1,\cdots,\tau(k)-1$,
         do (else, if $\tau(k;\omega)=0$, then go to step~3):
        \begin{align}
        & \mathrm{Update\, the\, inner\, variables}\, x_j(k,s),j=1,\cdots,N, \mathrm{by}: \nonumber\\
        \label{eqn-randomized-AL-update-primal}
        &         x_j(k,s+1) =\\
        \nonumber
        & \left\{\begin{array}{ll}
        \mathrm{argmin}_{x_j \in {\mathbb R}^d}
         (f_j(x_j)
         + \left(\mu_j(k) - \rho \overline{x}_j(k,s) \right)^\top x_j + \frac{\rho \|x_j\|^2}{2}) \\
         \mbox{$j={\hat{\imath}}(k,s)$} \\
          x_j(k,s+1) = x_j(k,s)\\
            \mbox{ else.}
       \end{array} \right. \\
         & \mathrm{Update\, the\, variables}\, \overline{x}_j(k,s), j=1,\cdots,N, \mathrm{by:} \nonumber \\
        & \overline{x}_j(k,s+1)\!\! =\!\! \left\{\!\! \begin{array}{ll}
        \!\!\sum_{l \in \Omega_j} \!\!\!\!W_{jl} x_l(k,s+1)
         &\!\!\!\!\!\!\mbox{$j\in {O_i}\!\!:i\!\!=\!\!{\hat{\imath}}(k,s)$} \\
          \overline{x}_j(k,s+1) = \overline{x}_j(k,s) &\!\!\!\!\!\!\mbox{else;} \end{array} \right.
          \end{align}
          and all nodes $j=1,\cdots,N$ set $x_j(k+1):={x}_j(k,s=\tau(k))$, $\overline{x}_j(k+1)=x_j(k,s=\tau(k))$.
        \STATE (\textbf{Outer iteration}) All nodes $j$ update the dual variables $\mu_j(k)$ via:
        \begin{eqnarray}
             \label{eqn-AL-algorithm-3-randomized}
             \mu_j(k+1) =  \mu_j(k) + \alpha \,\left(x_j(k+1) - \overline{x}_j(k+1) \right).
        \end{eqnarray}
        \STATE Set $k \mapsto k+1$ and go to step 2.
        }
        }
\end{algorithmic}
\end{algorithm}
For all $i$, and arbitrary fixed $k,s,$ Algorithm~3 defines $x_i(k,s) = x_i(k,s;\omega)$
for any outcome $\omega \in \cup_{t=s}^{\infty} \mathcal{A}_{k,t}.$ We formally
define $x_i(k,s;\omega)=0$, for any $\omega \in \Theta$, $\omega \notin \cup_{t=s}^{\infty} \mathcal{A}_{k,t}$.
 Thus, the random variable $x_i(k,s)$ is defined as in Algorithm~3 for $\omega \in \cup_{t=s}^{\infty} \mathcal{A}_{k,t}$,
 and $x_i(k,s;\omega)=0$, for $\omega \notin \cup_{t=s}^{\infty} \mathcal{A}_{k,t}.$
%
%
%
%

\textbf{Gradient primal updates}.
This algorithm variant is the same as Algorithm~3,
except that step~\eqref{eqn-randomized-AL-update-primal} is replaced by the following:
\begin{align}
\nonumber
&x_j(k,s+1)\!\!\! = \!\!\!\\
& \left\{\!\!\! \begin{array}{lll}
         \left( 1-\beta\rho \right)\,x_j(k,s) + \beta\rho \overline{x}_j(k,s)
         -\beta\left(\mu_j(k)+ \nabla f_j(x_j(k,s))\right)&\,\\
          \mbox{for $j\!\!=\!\!{\hat{\imath}}(k,s)$} \\
          x_j(k,s+1) = x_j(k,s)\\
          \mbox{ else.}
       \end{array} \right.
       \label{eqn-update-randomized-gradient}
\end{align}
Here, $\beta>0$ is the (primal) step-size parameter.

\section{Analysis of inexact augmented Lagrangian methods}
In this Section, we introduce our framework for the analysis of inexact AL algorithms~\eqref{eqn-update-primal}--\eqref{eqn-update-dual}.
 Subsection~{III-A} states our result, while Subsection~{III-B} proves the result through several auxiliary Lemmas.
 In Section~{IV}, we apply these results to each of the four distributed algorithms.
\subsection{Inexact AL algorithm: Convergence rate}
We consider an inexact version of algorithm~\eqref{eqn-update-primal}--\eqref{eqn-update-dual}.
Introduce compact notation, and denote by $x(k):=(x_1(k)^\top,...,x_N(k)^\top)^\top$,
and $\mu(k):=(\mu_1(k)^\top,...,\mu_N(k)^\top)^\top$. Recall the AL function in~\eqref{eqn-augmented-Lagrangian}.
For any $\mu \in {\mathbb R}^{N d}$, denote by $x^\prime(\mu):=\mathrm{arg\,min}_{x \in {\mathbb R}^{d N}} L_a(x;\mu)$.
The latter quantity is well-defined as the function $L_a(\cdot;\mu)$ is strongly convex in $x$, for any $\mu$.
Recall the weighted Laplacian matrix $\mathcal L=I-W$.
  We consider the following inexact AL method
 that updates the primal variable $x(k)$ and the dual variable $\mu(k)$ over iterations $k=0,1,...$.
 The primal initialization $x(0)$ is arbitrary, and the dual $\mu(0)=0$.
 For $k=0,1,...$, given $x(k)$, $\mu(k)$, perform the following update:
\begin{align}
\label{eqn-update-primal-NEW}
& x(k+1)\,\mathrm{be\,any\,point\,such\,that:\,\,}\\
& \|x(k+1)-x^\prime(\mu(k))\| \leq \xi\,\|x(k)-x^\prime(\mu(k))\| \nonumber\\
\label{eqn-update-dual-NEW}
& \mu(k+1) = \mu(k) + \alpha \,\left( \mathcal L \otimes I\right)x(k+1).
\end{align}
Update~\eqref{eqn-update-dual-NEW} is~\eqref{eqn-update-dual} rewritten in a compact form.
(Here $\mathcal L \otimes I$ is the Kronecker product of $\mathcal L$ and the $d \times d$ identity matrix.)
In~\eqref{eqn-update-primal-NEW}, the constant $\xi \in (0,1)$.
 Update~\eqref{eqn-update-primal-NEW} is an inexact version of~\eqref{eqn-update-primal}.
 Note that $x^\prime(\mu(k))$ corresponds to the exact AL update. We require
 that $x(k+1)$ be close to $x^\prime(\mu(k))$;
 more precisely, $x(k+1)$ be $\xi$ times closer to $x^\prime(\mu(k))$ than~$x(k)$.
  The motivation for this condition is the following.
  Given $\mu(k)$, we seek the new primal variable (ideally $x^\prime(\mu(k))$)
   via an iterative method, initialized by the previous primal variable $x(k)$.
   We stop the iterative method as soon as~\eqref{eqn-update-primal-NEW} is fulfilled.\footnote{As we will see in Section~{IV}, with our distributed
methods we do
   not verify the termination condition in~\eqref{eqn-update-primal-NEW} on-the-fly. Instead,
   given a desired $\xi$ and the network and function parameters, we set beforehand the number of
   inner iterations~$\tau$ such that~\eqref{eqn-update-primal-NEW} is automatically fulfilled.}

We now present our generic Theorem on~\eqref{eqn-update-primal-NEW}--\eqref{eqn-update-dual-NEW}.
We apply it on the four distributed methods in Section~{IV}.
Denote by $D_x:=\|x_1(0)-x^\star\|$, and
$D_{\mu}:=\left( \frac{1}{N}\sum_{i=1}^N \|\nabla f_i(x^\star)\|^2\right)^{1/2}$.

 \begin{theorem}
 \label{theorem-convergence-rate}
 Consider algorithm~\eqref{eqn-update-primal-NEW}--\eqref{eqn-update-dual-NEW}, and let Assumptions~\ref{assumption-bdd-hessian} and~\ref{assumption-network} hold.
 Further, let the algorithm and network parameters satisfy:
 \begin{align}
 \alpha &\leq h_{\mathrm{min}}+\rho\:\:\mathrm{and}\:\:
 \label{eqn-theorem-condition-linear-convergence}
 {{\xi}} < \frac{1}{3}\,\frac{\lambda_2(\mathcal L)\,h_{\mathrm{min}}}{\rho+h_{\mathrm{max}}}.
 \end{align}
  Then, at any node $i$,
  $x_i(k)$ generated by~\eqref{eqn-update-primal-NEW}--\eqref{eqn-update-dual-NEW} converges linearly to the solution $x^\star$
   of~\eqref{eqn-opt-prob-original}, with convergence factor:
  \begin{align}
  r := \max\left\{ \frac{1}{2}+\frac{3}{2}{{\xi}},
    \label{eqn-theorem-rate}
   \left( 1-\frac{\alpha\lambda_2(\mathcal{L})}{\rho+h_{\mathrm{max}}}\right)
  + \frac{3 \alpha}{h_{\mathrm{min}}} {{\xi}}   \right\}<1.
  \end{align}
  It holds:
  \begin{equation}
  \label{eqn-theorem-bound-Jacobi}
  \|x_i(k)-x^\star\| \leq r^{k}
  \sqrt{N} \max \left\{ {D_x},\frac{2{D_{\mu}}}{\sqrt{\lambda_2(\mathcal L)}h_{\mathrm{min}}} \right\}.
  \end{equation}
 \end{theorem}
Theorem~1 establishes that the inexact AL method converges to the \emph{primal} solution at the globally linear rate
in the number of \emph{outer} iterations, provided that $\xi$ is sufficiently small, and it
 quantifies the achieved rate as well as how small $\xi$ should be.
  We emphasize the interesting
  effect of constant~$D_{\mu}\!\!:=\!\!\left(\frac{1}{N}\! \sum_{i=1}^N \!\|\nabla f_i(x^\star)\|^2\!\right)^{1/2}\!\!$.
      It measures how difficult it is to solve~\eqref{eqn-opt-prob-original} by distributed methods like
       \eqref{eqn-update-primal}--\eqref{eqn-update-dual}--the larger,
       the more difficult the problem is. If, at an extreme,
       the $f_i$'s all have the same minimizer, say $y^\star$,
       then $y^\star$
      is also the minimizer of~\eqref{eqn-opt-prob-original} ($y^\star=x^\star$.)
      Such problem is ``easy,'' because nodes do not need to communicate with others to obtain
      the global solution to~\eqref{eqn-opt-prob-original}--``easyness'' of the problem agrees with the value $D_{\mu}=0$. On the other hand,
      if the local minimizers (of the $f_i$'s), say $y_i^\star$'s,
      are very different, then they may be very different from~$x^\star$.
      Hence, node $i$ needs to communicate with others to recover $x^\star$. This agrees with $D_{\mu}$ large in such scenarios. (See Lemma~\ref{lemma-solving-saddle-point-solves-original}
      that relates $D_{\mu}$ to the dual optimum.)

\subsection{Auxiliary results and proof of Theorem~1}
We now prove Theorem~1 by introducing several auxiliary objects and results.
 We
base our analysis on the following nonlinear saddle point system of equations:
{
\allowdisplaybreaks
\begin{align}
\label{eqn-saddle-point-system-1}
&\nabla F(x) + \mu + \rho \,(\mathcal{L}\otimes I) \,x =0 \\
\label{eqn-saddle-point-system-2}
& (\mathcal{L}\otimes I) x =0\\
\label{eqn-saddle-point-system-3}
& (1\otimes I)^\top \mu =0.
\end{align}
}
In~\eqref{eqn-saddle-point-system-1}, $\rho \geq 0$
 is the AL penalty parameter, and
$F: \mathbb R^{N \,d} \mapsto \mathbb R$ is
defined by $F(x)=F(x_1,\cdots,x_N)=f_1(x_1)+f_2(x_2)+\cdots+f_N(x_N)$.
In~\eqref{eqn-saddle-point-system-1},
$x, \mu \in {\mathbb R}^{N\,d}$ are the primal and dual variables, whose $i$-th coordinates correspond to
node $i$'s primal and dual variables, respectively.
In~\eqref{eqn-saddle-point-system-1}--\eqref{eqn-saddle-point-system-3} and in subsequent
text, Kronecker products $a \otimes b$ are always such that the left object $a$ is of size either $N \times 1$ or $N \times N$, while the right object is of size $d\times 1$ or $d \times d$. Henceforth, to simplify notation, we do not designate the objects' dimensions.
The next Lemma shows that
solving~\eqref{eqn-saddle-point-system-1}
solves~\eqref{eqn-opt-prob-original} at each node~$i$.
 \begin{lemma}
 \label{lemma-solving-saddle-point-solves-original}
 Consider optimization problem~\eqref{eqn-opt-prob-original} and the nonlinear system~\eqref{eqn-saddle-point-system-1}, and let Assumptions~\ref{assumption-bdd-hessian} and~\ref{assumption-network} hold. Then, there exists unique $(x^\bullet,\mu^\bullet)
 \in {\mathbb R}^{Nd} \times {\mathbb R}^{Nd}$ that satisfies~\eqref{eqn-saddle-point-system-1}--\eqref{eqn-saddle-point-system-3},
 with $x^\bullet = 1 \otimes x^\star$,
 where $x^\star$ is the solution to~\eqref{eqn-opt-prob-original}
 and $\mu^\bullet = - \nabla F(x^\star\,1)$.
 \end{lemma}
\begin{IEEEproof}
First show
$x^\bullet = 1 \otimes x^\star$ and $\mu^\bullet=-\nabla F(1 \otimes x^\star)$
 solves~\eqref{eqn-saddle-point-system-1}--\eqref{eqn-saddle-point-system-3}.
 Consider~\eqref{eqn-saddle-point-system-2}. We have
 $(\mathcal L \otimes I)x^\bullet = (\mathcal L \otimes I) (1 \otimes x^\star) =
 (\mathcal L \otimes 1)(I \otimes x^\star)=0$, since~$1$ is the unique eigenvector with eigenvalue~$0$ of the Laplacian
  for a connected network.
   Next:
   \[
   (1 \otimes I)^\top \mu^\bullet = -\sum_{i=1}^N \nabla f_i(x^\star)=0.
   \]
   The right equality holds because $x^\star$ is the solution to~\eqref{eqn-opt-prob-original}.
    Finally, because $(\mathcal L \otimes I)x^\bullet=0$ (already shown)
     and $\nabla F(x^\bullet)=-\mu^\bullet$, we have
      $(x^\bullet=1\otimes x^\star,\,\mu^\bullet=-\nabla F(1\otimes x^\star))$
      satisfy~\eqref{eqn-saddle-point-system-1}--\eqref{eqn-saddle-point-system-3}.
      The uniqueness is by the uniqueness of the solution to~\eqref{eqn-opt-prob-original}
      due to strong convexity.
%
\end{IEEEproof}
Next, introduce the following maps
$\Phi: {\mathbb R}^{Nd} \mapsto {\mathbb R}^{Nd}$, $\Psi: {\mathbb R}^{Nd} \mapsto {\mathbb R}^{Nd}$,
and $\Phi_i: {\mathbb R}^d \mapsto {\mathbb R}^d$, $i=1,...,N$:
\begin{eqnarray}
\label{eqn-Phi-map}
\Phi(x) &:=& \nabla F(x) + \rho\,I\,x\\
\label{eqn-Psi-map}
\Psi(x) &:=& \nabla F(x) + \rho\,\mathcal L \,x\\
\label{eqn-Phi-map-i}
\Phi_i(x) &:=& \nabla f_i(x) + \rho\,x.
\end{eqnarray}
Further, define the maps: $\Phi^{-1}:\,{\mathbb R}^{Nd} \rightarrow {\mathbb R}^{Nd}$, $\Psi^{-1}:\,{\mathbb R}^{Nd} \rightarrow {\mathbb R}^{Nd}$,
and $\Phi_i^{-1}:\,{\mathbb R}^d \rightarrow {\mathbb R}^d$ by:
%
%
\begin{align}
\label{eqn-Phi-map-inverse}
&\Phi^{-1}(\mu) \!\!:=\! \mathrm{argmin}_{y \in {\mathbb R}^{Nd}}\!\! \left( F(y) -\mu^\top y + \frac{\rho}{2}\|y\|^2\right)\\
\label{eqn-Psi-map-inverse}
&\Psi^{-1}(\mu) \!\!:=\! \mathrm{argmin}_{y \in {\mathbb R}^{Nd}} \!\!\left( F(y) -\mu^\top y + \frac{\rho}{2}y^\top \mathcal L y\right)\!\!\\
\label{eqn-Phi-map-inverse-i}
&\Phi^{-1}_i(\mu) := \mathrm{arg\,min\,}_{y \in {\mathbb R}^d} \left( f_i(y) -\mu_i^\top\, y + \frac{\rho}{2}\,\|y\|^2\right).
\end{align}
%
%
The cost function in~\eqref{eqn-Psi-map-inverse}
is precisely $L_a$ in~\eqref{eqn-augmented-Lagrangian}.
For any $\mu \in {\mathbb R}^{Nd}$, these maps are well-defined by Assumption~\ref{assumption-bdd-hessian} (This assumption ensures that there exists a unique solution
 in the minimizations in~\eqref{eqn-Phi-map-inverse} and~\eqref{eqn-Psi-map-inverse},
 as the costs in~\eqref{eqn-Phi-map-inverse} and~\eqref{eqn-Psi-map-inverse} are strongly convex.) Next, we have:
\[
\nabla F(\Phi^{-1}(\mu)) +\rho \,I \Phi^{-1}(\mu)=\mu=\Phi(\Phi^{-1}(\mu)),
\]
where the left equality is by the first order optimality conditions, from~\eqref{eqn-Phi-map-inverse}, and the right equality is by definition of $\Phi$ in~\eqref{eqn-Phi-map}.
Thus, the map $\Phi^{-1}$ is the inverse of $\Phi.$
 Likewise, the map $\Psi^{-1}$ is the inverse of $\Psi.$
By the inverse function theorem, e.g.,~\cite{CalculusBook},
the maps $\Phi^{-1}:\,{\mathbb R}^{Nd} \rightarrow {\mathbb R}^{Nd}$ and $\Psi^{-1}:\,{\mathbb R}^{Nd} \rightarrow {\mathbb R}^{Nd}$
 are continuously differentiable, with derivatives:
 \begin{eqnarray}
 \label{eqn-Phi_norms}
 \nabla \Phi^{-1}(\mu)  &=& \left( \,\nabla^2 F(\Phi^{-1}(\mu))+\rho\,I \, \right)^{-1}\\
 \label{eqn-Psi-norms}
 \nabla  \Psi^{-1}(\mu)  &=& \left(\, \nabla^2 F(\Psi^{-1}(\mu))+\rho\,(\mathcal L \otimes I)\, \right)^{-1}\\
 \label{eqn-Phi_norms-i}
 \nabla \Phi^{-1}_i(\mu)  &=& \left( \,\nabla^2 f_i(\Phi^{-1}_i(\mu))+\rho\,I \right)^{-1}.
 \end{eqnarray}
Note that invertibility is assured because $\nabla^2 F(x)$ and $\nabla^2 f_i(x_i)$
 are positive definite, $\forall x \in {\mathbb R}^{Nd}$, $\forall x_i \in {\mathbb R}^d$, and so are
 the matrices in~\eqref{eqn-Phi_norms}--\eqref{eqn-Phi_norms-i}.
Using the following identity for a continuously
differentiable map $h:\,{\mathbb R}^{Nd} \rightarrow {\mathbb R}^{Nd}$, $\forall u,v \in {\mathbb R}^{Nd}$:
\begin{equation}
\label{eqn-norm-3}
h(u)-h(v)\!\! =\!\! \left[\int_{0}^1 \nabla h(v+z(u-v))d z\right]\!\!(u-v),
\end{equation}
we obtain the following useful relations:
{\small{
\allowdisplaybreaks
\begin{align}
\label{eqn-Phi-inverse-useful}
&\Phi^{-1}(\mu_1) - \Phi^{-1}(\mu_2) = R_{\Phi}(\mu_1,\mu_2)\,(\mu_1-\mu_2), \\
& R_{\Phi}(\mu_1,\mu_2)\!\! :=\!\! \int_{z=0}^1\!\!\! \!\! \nabla \Phi^{-1}(\mu_1+z(\mu_2-\mu_1))\,d z \nonumber \\
\label{eqn-Psi-inverse-useful}
&\Psi^{-1}(\mu_1) - \Psi^{-1}(\mu_2) = R_{\Psi}(\mu_1,\mu_2)(\mu_1-\mu_2) ,\\
& R_{\Psi}(\mu_1,\mu_2)\!\! :=\!\!\int_{z=0}^1\!\!  \nabla \Psi^{-1}(\mu_1+z(\mu_2-\mu_1))\,d z \nonumber \\
\label{eqn-Phi-inverse-useful-i}
&\Phi^{-1}_i(\mu_1) - \Phi^{-1}_i(\mu_2) = R_{\Phi,i}(\mu_1,\mu_2)\,(\mu_1-\mu_2),\\
& R_{\Phi,i}(\mu_1,\mu_2) \!\!:=\!\!\!\!\int_{z=0}^1\!\!\!\!\!\!  \nabla \Phi^{-1}_i(\mu_1+z(\mu_2-\mu_1))d z. \nonumber
\end{align}
}}
By Assumption~\ref{assumption-bdd-hessian}:
 $h_{\mathrm{min}}\,I \preceq \nabla^2 F(x) \preceq h_{\mathrm{max}} \,I$, $\forall x \in {\mathbb R}^{Nd}.$
Using the latter, \eqref{eqn-Phi_norms}, \eqref{eqn-Psi-norms}, \eqref{eqn-norm-3},
and $\mathcal L=I-W$, $0 \preceq \mathcal L \preceq I$ ($W \succ 0$, symmetric, stochastic),
we obtain the following properties of the $(Nd) \times (Nd)$ matrices $R_{\Phi}(\mu_1,\mu_2)$ and $R_{\Psi}(\mu_1,\mu_2)$, and
$d \times d$ matrices $R_{\Phi,i}(\mu_1,\mu_2)$:
\begin{eqnarray}
\label{eqn-Phi-inv-bounds}
\frac{1}{h_{\mathrm{max}}+\rho} \,I &\preceq& R_{\Phi}(\mu_1,\mu_2)  \preceq \frac{1}{h_{\mathrm{min}}+\rho}\,I,\\
&\,&\forall \mu_1,\mu_2\in {\mathbb R}^{Nd} \nonumber \\
\label{eqn-psi-inv-bounds}
\frac{1}{h_{\mathrm{max}}+\rho}\,I &\preceq& R_{\Psi}(\mu_1,\mu_2) \preceq
\left(h_{\mathrm{min}}I+\rho(\mathcal L\otimes I)   \right)^{-1},\\
&\,&\forall \mu_1,\mu_2\in {\mathbb R}^{Nd} \nonumber\\
\label{eqn-phi-inv-bounds-i}
\frac{1}{h_{\mathrm{max}}+\rho}I \!\! &\preceq& \!\! R_{\Phi,i}(\mu_1,\mu_2) \!\! \preceq \!\! \frac{1}{h_{\mathrm{min}}+\rho}I,\\
&\,&\forall \mu_1,\mu_2 \!\in \!{\mathbb R},\:\forall \mu_1,\mu_2\in {\mathbb R}^{d}. \nonumber
\end{eqnarray}
The right inequality in~\eqref{eqn-psi-inv-bounds} holds because, $\forall \mu$,
$\nabla^2 F(\Psi^{-1}(\mu))+\rho\,(\mathcal L \otimes I)$
 $\succeq h_{\mathrm{min}}I+\rho\,(\mathcal L \otimes I)$ (due to Assumption~1), and
 so $[\,\nabla^2 F(\Psi^{-1}(\mu))+\rho\,(\mathcal L \otimes I)\,]^{-1}$
  $\preceq [\, h_{\mathrm{min}}I+\rho\,(\mathcal L \otimes I) \,]^{-1}$. Denote by $\widetilde{x}(k):=x(k)-x^\bullet$
 and $\widetilde{\mu}(k):=\mu(k)-\mu^\bullet$
 the primal and dual errors, respectively. Also,
 write $x^\prime(k):=x^\prime(\mu(k))$, to simplify notation. We now state and prove several Lemmas
 that allow us to prove Theorem~1. We prove these lemmas
 assuming $d=1$, to avoid further extensive use of Kronecker products; the proofs extend
 to generic $d>1.$
 We first upper bound the primal error $\|\widetilde{x}(k+1)\|.$
 \begin{lemma}[Primal error]
 \label{lemma-primal-bound}
 Let Assumptions~\ref{assumption-bdd-hessian}, \ref{assumption-network} hold. Then,
 for $k=0,1,\cdots$
 \begin{align*}
 \|\widetilde{x}(k+1)\| \leq {\xi}
 \|\widetilde{x}(k)\| 
 + \frac{1}{h_{\mathrm{min}}} \left( 1+ {\xi} \right) \|\widetilde{\mu}(k)\|.
 \end{align*}
 \end{lemma}
 \begin{IEEEproof}
 Write $\widetilde{x}(k+1)=(x(k+1)-x^\prime(k+1)) + (x^\prime(k+1)-x^\bullet)$.
  Then, $\|\widetilde{x}(k+1)\| \leq \|x(k+1)-x^\prime(k+1)\| + \|x^\prime(k+1)-x^\bullet\|$.
  From~\eqref{eqn-update-primal-NEW}, we know that
  $\|x(k+1)-x^\prime(k+1)\| \leq \xi \|x(k)-x^\prime(k+1)\|$.
  The latter is further upper bounded as:
  $
  \|x(k+1)-x^\prime(k+1)\|$ $\leq \xi \|x(k)-x^\bullet + x^\bullet - x^\prime(k+1)\| $ $\leq \xi \|\widetilde{x}(k)\|+\xi\|x^\bullet - x^\prime(k+1)\|.
  $
  Hence,
  \begin{equation}
  \label{eqn-11111}
  \|\widetilde{x}(k+1)\| \leq \xi \|\widetilde{x}(k)\| + (1+\xi) \|x^\prime(k+1)-x^\bullet\|.
  \end{equation}
  It remains to upper bound $\|x^\prime(k+1)-x^\bullet\|$.
   Note that $x^\bullet = \Psi^{-1}(-\mu^\bullet)$. Using the latter and~\eqref{eqn-Psi-inverse-useful}, we obtain:
  \begin{align}
  &x^\prime(k+1)-x^\bullet = \Psi^{-1}(-\mu(k)) - \Psi^{-1}(\mu^\bullet) \nonumber \\
  \label{eqn-x-prime-x-bullet}
  &= -R_{\Psi}(k)\,(\mu(k)-\mu^\bullet),
  \end{align}
  with $R_{\Psi}(k):=R_{\Psi}(-\mu(k),-\mu^\bullet)$.
  This, with~\eqref{eqn-psi-inv-bounds}, and $\widetilde{\mu}(k)=\mu(k)-\mu^\bullet$, gives:
  \begin{equation}
  \label{eqn-apply-al-proofs}
  \|x^\prime(k+1)-x^\bullet\| \leq \frac{1}{h_{\mathrm{min}}}\,\|\widetilde{\mu}(k)\|.
  \end{equation}
The result follows from~\eqref{eqn-11111} and~\eqref{eqn-apply-al-proofs}.
 \end{IEEEproof}
 %
 %
 %

Since our final goal is to bound the primal error, rather than bounding $\widetilde{\mu}(k)=\mu(k)-\mu^\bullet$, it turns out to be
more useful to bound a certain transformed quantity.
Represent the weighted Laplacian matrix $\mathcal L $ through its (reduced) eigen-decomposition (we do not include the pair $\left(0,q_1\right)$)
$\mathcal L = Q \widehat{\Lambda} Q^\top=\sum_{i=2}^N \lambda_i\, q_i q_i^\top$, where $\left(\lambda_i, q_i\right)$ is the $i$-th eigenvalue, eigenvector pair ($\lambda_i>0$, for all $i=2,\cdots,N$);  $Q=[q_2,\cdots,q_N]$; and  $\widehat{\Lambda}=\diag\left[\lambda_2,\cdots,\lambda_N\right]$.
  Instead of bounding the dual error, we bound the norm of $\widetilde{\mu}^{\prime \prime}(k)\in {\mathbb R}^{N-1}$ that we define:
   \begin{equation}
   \label{eqn-mu-prime}
   \widetilde{\mu}^\prime(k)\!\!:=\!\!Q^\top \widetilde{\mu}(k) \!\!\in\! {\mathbb R}^{N-1}\:\mathrm{and}\:
   \widetilde{\mu}^{\prime \prime}(k)\!\!:=\!\!\widehat{\Lambda}^{-1/2} \widetilde{\mu}^\prime(k).
   \end{equation}
  %
  %
  %
\begin{lemma}[Dual error]
\label{lemma-dual-error}
Let $\alpha \leq h_{\mathrm{min}}+\rho$, and let Assumptions~\ref{assumption-bdd-hessian} and~\ref{assumption-network} hold. Then,
for all $k=0,1,\cdots$
\begin{align*}
\|\widetilde{\mu}^{\prime \prime}(k+1)\|
\!\leq \!
\left[\!\left(\!\! 1\!- \!\frac{\alpha\lambda_2(\mathcal L)}{h_{\mathrm{max}}+\rho}\!\!\right)
\!\!+ \!\!\frac{\alpha}{h_{\mathrm{min}}}{\xi} \right] 
\|\widetilde{\mu}^{\prime \prime}(k)\|
+ \alpha {{\xi}} \|\widetilde{x}(k)\|.
\end{align*}
\end{lemma}
\begin{IEEEproof}
%
 Because $\mathcal L x^\bullet=\mathcal L x^\star\,1=0$:
\[
\mathcal L x(k+1) = \mathcal L (x(k+1)-x^\prime(k+1))+\mathcal L(x^\prime(k+1)-x^\bullet).
\]
Using this and subtracting $\mu^\bullet$
 from both sides of~\eqref{eqn-update-dual-NEW}:
 \begin{align}
 & \widetilde{\mu}(k+1)
 =
 \widetilde{\mu}(k) + \alpha \,\mathcal L (x^\prime(k+1)-x^\bullet) \\
 & + \alpha\,\mathcal{L} (x(k+1)-x^\prime(k+1)). \nonumber
 \end{align}
Further, using~\eqref{eqn-x-prime-x-bullet}, we get:
\begin{equation}
\label{eqn-dual-proof-AL-1}
\!\widetilde{\mu}(\!k\!+\!1\!)\!\!
 = \!\!\left(\!\!I \!\!-\! \!\alpha\!\mathcal LR_{\Psi}\!(k)\!\! \right)\!
 \widetilde{\mu}(k)\! + \!\alpha\mathcal L \!(x(k\!+\!1)\!-\!x^\prime(k\!+\!1))\!.\!\!
\end{equation}
Now, recall $\widetilde{\mu}^\prime(k)$ in~\eqref{eqn-mu-prime}.
It is easy to see that:
\begin{eqnarray}
\label{eqn-dual-proof-AL-2}
\|\widetilde{\mu}^\prime(k)\|\!\!=\!\!\|\widetilde{\mu}(k)\| ,\,\:\:\:Q Q^\top \widetilde{\mu}(k)=\widetilde{\mu}(k).
\end{eqnarray}
Indeed, note that $1^\top \mu(k) = 1^\top \mu(k-1)+\alpha 1^\top \mathcal L x(k)
= 1^\top \mu(k-1)=\cdots=1^\top \mu(0)=0$, because $\mu(0)=0$ (by assumption.)
 Also, $1^\top \mu^\bullet = 0$ (see Lemma~\ref{lemma-solving-saddle-point-solves-original}.)
  Therefore, $1^\top \widetilde{\mu}(k)=0$, $\forall k$.
   Now, as $q_1=\frac{1}{\sqrt{N}}\,1$, we have
   $Q Q^\top \widetilde{\mu}(k)=\sum_{i=2}^N q_i q_i^\top \widetilde{\mu}(k)=
   \sum_{i=1}^N q_i q_i^\top \widetilde{\mu}(k)=\widetilde{\mu}(k)$; thus,
   the second equality in~\eqref{eqn-dual-proof-AL-2}.
   For the first equality in~\eqref{eqn-dual-proof-AL-2},
   observe that: $\|\widetilde{\mu}^\prime(k)\|^2
    = (\widetilde{\mu}^\prime(k))^\top \widetilde{\mu}^\prime(k)
    = \widetilde{\mu}(k)^\top Q Q^\top \widetilde{\mu}(k)=\|\widetilde{\mu}(k)\|^2$.

Next, multiplying~\eqref{eqn-dual-proof-AL-1} from the left by $Q^\top$,
expressing $\mathcal L = Q \widehat{\Lambda} Q^\top$, and using~\eqref{eqn-dual-proof-AL-2}, obtain:
\begin{align}
& \widetilde{\mu}^\prime(k+1) = \left(\,I - \alpha\,\widehat{\Lambda}\,Q^\top R_{\Psi}(k)\,Q  \,\right)\,\mu^\prime(k) \nonumber\\
\label{eqn-mu-prime-AL-proof-new}
& +\alpha\,\widehat{\Lambda}\,Q^\top \,(x(k+1)-x^\prime(k+1)).
\end{align}
Further, recall $\widetilde{\mu}^{\prime \prime}(k)$ in~\eqref{eqn-mu-prime}.
Multiplying~\eqref{eqn-mu-prime-AL-proof-new}
from the left by $\widehat{\Lambda}^{-1/2}$, we obtain:
\begin{align}
\nonumber
& \widetilde{\mu}^{\prime \prime}(k+1) = \left(\,I - \alpha\,\widehat{\Lambda}^{1/2}\,Q^\top R_{\Psi}(k)\,Q \widehat{\Lambda}^{1/2} \,\right)\,\widetilde{\mu}^{\prime \prime}(k)\\
\label{eqn-mu-prime-AL-proof-new-2}
& +\alpha\,\widehat{\Lambda}^{1/2}\,Q^\top \,(x(k+1)-x^\prime(k+1)).
\end{align}
Next, using variational characterizations of minimal and maximal eigenvalues, we can verify:
\begin{equation}
\label{eqn-matrica-vece-manje-AL-proofs}
\frac{\lambda_2}{h_{\mathrm{max}}+\rho}\,I \preceq \widehat{\Lambda}^{1/2}\,Q^\top R_{\Psi}(k)\,Q \widehat{\Lambda}^{1/2}
 \preceq
 \frac{1}{h_{\mathrm{min}}+\rho}\,I.
\end{equation}
The right inequality in~\eqref{eqn-matrica-vece-manje-AL-proofs} holds because of the following.
 First, use the right inequality in~\eqref{eqn-psi-inv-bounds} to show
 $\widehat{\Lambda}^{1/2}\,Q^\top R_{\Psi}(k)\,Q \widehat{\Lambda}^{1/2}
 \preceq \widehat{\Lambda}^{1/2}\,Q^\top [\,h_{\mathrm{min}}I+\rho \mathcal L  \,]^{-1}\,Q \widehat{\Lambda}^{1/2} $.
  (Note that $\widehat{\Lambda}$ is $(N-1)\times (N-1)$,
 $Q$ is $N \times (N-1)$, and $[\,h_{\mathrm{min}}I+\rho \mathcal L  \,]^{-1}$
  is $N \times N$.) Next, decompose the
 $N \times N$ matrix
 $[\,h_{\mathrm{min}}I+\rho \mathcal L  \,]^{-1}$ via the ($N \times N$) eigenvalue decomposition,
  and use orthogonality of the eigenvectors of~$\mathcal L$ to show that the ($(N-1)\times (N-1)$) matrix:
  $\widehat{\Lambda}^{1/2}\,Q^\top R_{\Psi}(k)\,Q \widehat{\Lambda}^{1/2} $
   $\preceq \widehat{\Lambda}^{1/2} [\,h_{\mathrm{min}}I+\rho \widehat{\Lambda}\,]^{-1}  \widehat{\Lambda}^{1/2}$. The maximal eigenvalue
   of $\widehat{\Lambda}^{1/2} [\,h_{\mathrm{min}}I+\rho \widehat{\Lambda}\,]^{-1}  \widehat{\Lambda}^{1/2}$
   is $\frac{1}{h_{\mathrm{min}}/\lambda_N(\mathcal L)+\rho} \leq \frac{1}{h_{\mathrm{min}}+\rho}.$
   Next, by Assumption, $\alpha \leq h_{\mathrm{min}}+\rho$,
and so:
\begin{equation}
\label{eqn-matrica-ineqiualties}
\|I - \alpha\,\widehat{\Lambda}^{1/2}\,Q^\top R_{\Psi}(k)\,Q \widehat{\Lambda}^{1/2}\| \leq 1-\frac{\alpha\,\lambda_2}{h_{\mathrm{max}}+\rho}.
\end{equation}
Using~\eqref{eqn-matrica-ineqiualties}, $\|\widehat{\Lambda}^{1/2}\|\leq 1$ (as $0 \preceq \mathcal L \preceq I$), $\|Q\|=1,$
 and Lemma~\ref{lemma-inexact-dual-ascent-x-x-prime}, we get:
 {\small{
\begin{align*}
& \|\widetilde{\mu}^{\prime \prime}(k+1)\|
\leq
\left(  1-\frac{\alpha\,\lambda_2}{h_{\mathrm{max}}+\rho} \right) \|\widetilde{\mu}^{\prime \prime}(k)\|\\
&
+ \alpha {{\xi}} \,\|\widetilde{x}(k)\|
+\alpha {{\xi}} \frac{\|\widetilde{\mu}(k)\|}{h_{\mathrm{min}}}.
\end{align*}}}
 Finally, using $\|\widetilde{\mu}(k)\|
 =\|\widetilde{\mu}^\prime(k)\|=
 \|{\widehat{\Lambda}}^{1/2}\widetilde{\mu}^{\prime \prime}(k)\|
 \leq
 \|\widetilde{\mu}^{\prime \prime}(k)\|,
 $
 we obtain the desired result.
\end{IEEEproof}

We are now ready to prove Theorem~\ref{theorem-convergence-rate}.

\begin{IEEEproof}[Proof of Theorem~\ref{theorem-convergence-rate}]
 Introduce $\nu(k):=\frac{2}{h_{\mathrm{min}}}\|\widetilde{\mu}(k)\|$. Further,
denote by $c_{11}:={{\xi}}$,
$c_{12}:=\frac{1}{2 } \left[\,1+{{\xi}}\,\right]$;
 $c_{21}:=\frac{2\,\alpha}{h_{\mathrm{min}}} {{\xi}}$,
 and $c_{22}:=\left(  1-\frac{\alpha\,\lambda_2}{h_{\mathrm{max}}+\rho} \right)+\frac{\alpha}{h_{\mathrm{min}}}
 {{\xi}}.$
  Using $\|\widetilde{\mu}(k)\| \leq \|\widetilde{\mu}^{\prime \prime}(k)\|$,
  Lemma~\ref{lemma-primal-bound}, and Lemma~\ref{lemma-dual-error}, we obtain:
  \begin{equation*}
  \max\left\{ \|\widetilde{x}(k+1)\|,\,\nu(k+1) \right\}
  \leq r\,\max\left\{  \|\widetilde{x}(k)\|,\,\nu(k) \right\},
  \end{equation*}
with $r=\max\left\{  c_{11}+c_{12},\,c_{21}+c_{22} \right\}.$
Unwinding the recursion, using
%
%
$
\|\widetilde{x}(k)\|\!\!\!\leq\!\!\! \max\{\|\widetilde{x}(k)\|,\,\nu(k)\},
$
%
$\nu(0)=\frac{2}{h_{\mathrm{min}}} \| \widehat{\Lambda}^{-1/2} Q^\top \widetilde{\mu}(0)\|=
\frac{2}{h_{\mathrm{min}}} \|\widehat{\Lambda}^{-1/2}\,Q^\top \,(-\nabla F(x^\star\,1))\|
\!\!\!\leq\!\!\! \frac{2}{h_{\mathrm{min}}\sqrt{\lambda_2}}\sqrt{N} {D_{\mu}}$, obtain~\eqref{eqn-theorem-bound-Jacobi}.

It remains to show that $r<1$ if conditions~\eqref{eqn-theorem-condition-linear-convergence} hold.
Note that:
$
c_{11} + c_{12} = \frac{1}{2} + \frac{3}{2}{\xi},
$
 and so $c_{11}+c_{12}<1$ if:
$
{\xi} < \frac{1}{3}.
$
Next, note that:
$
c_{21}+c_{22} = \left( 1-\frac{\alpha\,\lambda_2}{\rho+h_{\mathrm{max}}} \right) +\frac{3 \alpha}{h_{\mathrm{min}}}
{\xi},
$
and so $c_{21}+c_{22}<1$ if:
$
{\xi} < \frac{1}{3}\left( \frac{h_{\mathrm{min}} \lambda_2}{\rho+h_{\mathrm{max}}}\right).
$
Combining the last two conditions,
obtain $r<1$ if conditions~\eqref{eqn-theorem-condition-linear-convergence} hold.
The proof is complete.
\end{IEEEproof}

\section{Analysis of distributed augmented Lagrangian methods}
In this Section, we specialize our results from Section~{III} to each of the four distributed AL algorithm variants.
More precisely, we characterize the quantity~$\xi$ in~\eqref{eqn-update-primal-NEW} with each method.
 This, with Theorem~1, allows us to establish convergence rates in the inner iterations.

With each of the four variants, we use compact notation:
$x(k)=(x_1(k)^\top,...,x_N(k)^\top)^\top$,
$\mu(k)=(\mu_1(k)^\top,...,\mu_N(k)^\top)^\top$, and
$x(k,s)=(x_1(k,s)^\top,...,x_N(k,s)^\top)^\top$.
We start with the deterministic Jacobi variant.
\begin{lemma}[Deterministic Jacobi]
  \label{lemma-inexact-dual-ascent-x-x-prime}
Consider the distributed AL algorithm with deterministic Jacobi primal updates and $\tau$ inner iterations. Further, let Assumptions~\ref{assumption-bdd-hessian} and~\ref{assumption-network} hold.
   Then, for all $k=0,1,\cdots$:
  \begin{align*}
  \left\| x(k+1)-x^\prime(k+1)  \right\| \leq \left( \frac{\rho}{\rho+h_{\mathrm{min}}} \right)^\tau \,\left\| x(k)-x^\prime(k+1)  \right\|.
  \end{align*}
  \end{lemma}
  \begin{IEEEproof}
  Recall that $x^\prime(k+1)=\mathrm{arg\,min}_{x \in {\mathbb R}^N}L_a(x;\mu(k))$. From the
  corresponding first order optimality conditions, we have:
   $\nabla F(x^\prime(k+1))+\rho \,\mathcal L\,x^\prime(k+1)=-\mu(k)$. Hence, using $\mathcal L = I-W$ and the definition of $\Phi$ in~\eqref{eqn-Phi-map}:
  \begin{equation}
  \label{eqn-proof-1-AL-lemma}
  x^\prime(k+1) = \Phi^{-1} \left( \,\rho\,W x^\prime(k+1)-\mu(k)  \,\right).
  \end{equation}
  Fix $s$, $0\leq s\leq \tau-1$. Next, from Algorithm~1 and
  definition of $\Phi$:
  \begin{eqnarray}
         \label{eqn-AL-algorithm-1-compact}
         x(k,s+1) &=& \Phi^{-1} \left( \,  \rho\,W\,x(k,s)-\mu(k)  \,  \right);
  \end{eqnarray}
  Subtracting $x^\prime(k+1)$ from both sides of~\eqref{eqn-AL-algorithm-1-compact},
  and using~\eqref{eqn-proof-1-AL-lemma} and~\eqref{eqn-Phi-inverse-useful}:
  \begin{align*}
  x(k,s+1)\!-\!x^\prime(k+1)\!\! = \!\!R_{\Phi}(s)\rho W(x(k,s)\!-\!x^\prime(k+1)),
  \end{align*}
  where~$R_{\Phi}(s)\!\!:=\!\!R_{\Phi}\!\!\left(\!\rho W x(k,s)\!\!-\!\!\mu(k),\! \rho W x^\prime(k+1)\!\!-\!\!\mu(k)\!\right)$. Using~\eqref{eqn-Phi-inv-bounds} and $\|W\|=1$, obtain:
  \begin{align*}
  \|\!x(k,s\!+\!1)\!-\!x^\prime\!(k\!+\!1)\|\!\! \leq \!\! \left(\!\frac{\rho}{\rho+h_{\mathrm{min}}}\!\right)\!\!\|x(k,s)\!-\!x^\prime(\!k\!+\!1\!)\|.
  \end{align*}
  Applying this for $s=0,1,\cdots,\tau-1$,
   using $x(k,\tau)=x(k+1)$, $x(k,0)=x(k)$, get:
  {
  \allowdisplaybreaks
  {\small{
  \begin{align}
  &
  \|x(k\!+\!1)\!-\!x^\prime(k\!+\!1)\| \!\leq \!\left(\!\frac{\rho}{\rho+h_{\mathrm{min}}}\!\!\right)^{\!\!\tau}\hspace{-2mm} \|x(k)\!\!-\!\!x^\prime(k\!+\!1)\|.
  \end{align}
  }}}
  %
  \end{IEEEproof}

The immediate corollary of Lemma~5 is that, for the
distributed AL algorithm with Jacobi primal updates,
Theorem~1 holds with $\xi:=\left(\frac{\rho}{\rho+h_{\mathrm{min}}}\right)^\tau$.
In other words, if the conditions on the system parameters in Theorem~1 hold, the distributed AL algorithm converges linearly in the outer iterations.
  Furthermore, as the number of inner iterations is fixed and equals $\tau$,
  the algorithm also converges linearly in the number of inner iterations, and hence in
  the number of per-node communications, with the convergence factor~$r^{1/\tau}$.
  Note that, for any choice of $\rho \geq 0$, we can choose
  $\alpha$ and $\tau$ such that linear convergence is assured.
   Setting $\rho=h_{\mathrm{max}}$,
   $\alpha=h_{\mathrm{min}}+\rho$, and
   $
   \tau = \left\lceil  \frac{  \log(6 \gamma/\lambda_2)      }{ \log(1+1/\gamma)     }  \right\rceil,
   $
we obtain the convergence factor at outer iterations
$r=1-\Omega(\lambda_2)$. Hence, interestingly,
we can eliminate the negative effect of the condition number $\gamma$
 at the outer iterations level. Of course,
 we pay a price at the inner iterations level,
 where the convergence factor is
 $r^{1/\tau}=1-\Omega\left(  \frac{\lambda_2 \log(1+1/\gamma)} {\log(\gamma/\lambda_2)}\right)$.

We remark that, for a reasonable choice of the step-size $\alpha$ and the AL penalty $\rho$, e.g., $\alpha=\rho=h_{\mathrm{min}}$,
our results do not guarantee linear convergence for~$\tau=1$. (Hence, we do not guarantee convergence for $\tau=1$.) However, we know from the literature that, for any choice of $\alpha =\rho>0$, the algorithm with Jacobi updates and $\tau=1$ (distributed ADMM) converges globally linearly to the primal solution~\cite{ADMMYinJournal}. This, in particular, means that, for $\tau=1$,
 the algorithm converges at a globally linear rate if and only if it converges (at any rate).


We now consider the deterministic gradient variant.
\begin{lemma}[Deterministic gradient]
  \label{lemma-inexact-dual-ascent-x-x-prime-gradient}
  Consider the distributed AL algorithm with deterministic gradient primal updates with $\tau$ inner iterations
  and the primal step-size $\beta\leq 1/(h_{\mathrm{max}}+\rho)$. Further, let Assumptions~\ref{assumption-bdd-hessian} and~\ref{assumption-network} hold.
   Then, for all $k=0,1,\cdots$:
  \begin{align*}
  \left\| x(k+1)-x^\prime(k+1)  \right\| \leq \left( 1-\beta\,h_{\mathrm{min}} \right)^\tau \,\left\| x(k)-x^\prime(k+1)  \right\|.
  \end{align*}
  \end{lemma}
  %
%
%
%
%
%
%
%
\begin{IEEEproof}
Using $\mathcal L = I- W$ and compact notation, the update~\eqref{eqn-update-gradient-descent} is rewritten as:
\begin{equation}
\label{eqn-compact-notation-gradient-update}
\!x(k,\!s\!+\!1)\!\!=\!\!x(k,\!s)\!\!-\!\!\beta \!\left(\! \rho\mathcal{L} \!x(k,s)\!\! + \!\!\mu(k) \!\!+\!\!\nabla F(x(k,s)) \right)\!.\!\!
\end{equation}
This is the gradient descent on $L_a(\cdot;\mu(k))$ in~\eqref{eqn-augmented-Lagrangian}.
  As $x^\prime(k+1)$ satisfies  $\rho\,\mathcal L \,x^\prime(k+1) + \mu(k) +\nabla F(x^\prime(k+1))=0,$ we have:
\begin{equation}
\label{eqn-to-subtract-H-S}
\!x^\prime\!(\!k\!+\!1\!)\!\!=\!\!x^\prime\!\!(\!k\!+\!1\!)\!-\!\beta \!\!\left(\! \rho\!\mathcal{L} \!x^\prime(\!k\!+\!1\!) \!+\! \mu(k)\! +\!\nabla\! F\!(x^\prime(k\!+\!1)) \!\!\right)\!.\!\!
\end{equation}
%
%
%
Further, by Assumption~\ref{assumption-bdd-hessian},
$\nabla F:\,{\mathbb R}^N \rightarrow {\mathbb R}^N$
 is continuously differentiable, and it holds:
 \begin{eqnarray}
 &\,& \nabla F(x(k,s))-\nabla F(x^\prime(k+1)) = \nonumber\\
 &\,&
 \left[\int_{z=0}^1 \nabla^2 F \left(x^\prime(k+1)+z(x(k,s)-x^\prime(k+1))\right)d z\right] \nonumber\\
 &\times& (x(k,s)-x^\prime(k+1)) \nonumber \\
 \label{eqn-H-F-equality}
 &=:&
 H_F\left( s \right)(x(k,s)-x^\prime(k+1)).
 \end{eqnarray}
Further, by Assumption~\ref{assumption-bdd-hessian}, the matrix
$H_F\left( s \right)$ satisfies:
\begin{equation}
\label{eqn-H-S-bounds-upper-lower}
h_{\mathrm{min}}\,I \preceq H_F(s) \preceq h_{\mathrm{max}}\,I.
\end{equation}
Using~\eqref{eqn-H-F-equality},
and subtracting~\eqref{eqn-to-subtract-H-S} from~\eqref{eqn-compact-notation-gradient-update}, we obtain:
\begin{align}
& x(k,s+1)-x^\prime(k+1) = \left( I - \beta\,\rho\,\mathcal L -\beta\, H_F(s) \right) \nonumber \\
\label{eqn-apply-latter-gradient-proof-lemma}
&\times (x(k,s)-x^\prime(k+1)).
\end{align}
Consider the matrix $\left( I - \beta\,\rho\,\mathcal L -\beta\, H_F(s) \right)$.
As $\beta \leq \frac{1}{\rho+h_{\mathrm{max}}}$ (by assumption),
using \eqref{eqn-H-S-bounds-upper-lower} and $0 \preceq \mathcal L \preceq I$,
get:
 $\left( I - \beta\,\rho\,\mathcal L -\beta\, H_F(s) \right) \succeq 0.$
 Thus, $\|I - \beta\,\rho\,\mathcal L -\beta\, H_F(s)\|
 \leq 1-\lambda_1\left( \beta\,\rho\,\mathcal L + \beta\,H_F(s)  \right)
 \leq 1-\beta\,h_{\mathrm{min}}$.
 Applying this bound to~\eqref{eqn-apply-latter-gradient-proof-lemma},
 obtain the inequality:
 \begin{equation}
 \label{eqn-important-gradient-lemma}
 \|x(k,s\!+\!1)\!-\!x^\prime\!(k\!+\!1)\! \|\!\!\leq\!\! \left(\! 1\!-\!\beta \!h_{\mathrm{min}}\! \right)\! \|\!x(k,s)\!-\!x^\prime(k\!+\!1)\! \|\!.\!\!
 \end{equation}
%
%
%
Applying~\eqref{eqn-important-gradient-lemma}
 for $s\!=0,\cdots,\!\tau-1$,
  using $x(k,s\!=\!0)\!=\!x(k)$, and $x(k,s\!=\!\tau)\!=\!x(k\!+\!1)$,
   we obtain the desired result.
\end{IEEEproof}
The immediate corollary of Lemma~6 is that Theorem~1 holds
for the deterministic gradient variant, with $\xi=(1-\beta\,h_{\mathrm{min}})^\tau$.
 Hence, under conditions of Theorem~1,
 the algorithm converges linearly in the number of inner iterations,
 with the convergence factor~$r^{1/\tau}$. This implies the linear
 convergence both in the number of per-node communications and in
 the number of per-node gradient evaluations.
  Setting $\rho=h_{\mathrm{max}}$, $\beta=\frac{1}{h_{\mathrm{max}}+\rho}$,
  and:
  $
  \tau = \left\lceil   \frac{   \log(6 \gamma/\lambda_2)      }{  \log\left(1+\frac{1}{2\gamma-1}\right)    }    \right\rceil,
  $
gives the convergence factor in the inner iterations
as $r^{1/\tau}=1-\Omega\left( \frac{\lambda_2\log(1+1/\gamma)}{\log(\gamma/\lambda_2)}  \right).$

Note that, for reasonable choices of $\alpha,\beta$, and $\rho$, e.g.,
$\alpha=\rho=h_{\mathrm{min}}$, $\beta=1/(\rho+h_{\mathrm{max}})$, our results do not guarantee convergence nor
linear convergence rates when we set $\tau=1$. Reference~\cite{ControlApproach} establishes global convergence of a similar
 algorithm for $\tau=1$, $\rho=0$, and a sufficiently small~$\alpha$ and $\beta$.
  An interesting research direction is to explore whether there is a boundary between stability results and global linear rates.
 In other words, setting $\tau=1$, an open problem is whether for certain choices of $\alpha,\beta,$ and $\rho$ the algorithm converges at globally sub-linear rates. (Recall that this scenario does not occur with the Jacobi variant.) Another important open problem is
 to research whether, for $\tau=1$, there exists a choice of $\alpha,\beta$, and $\rho$ that ensures globally linear rates.
  Recall the random model in Subsection~{II-C} and the randomized Gauss-Seidel method.

 \begin{lemma}[Randomized Gauss-Seidel]
  \label{lemma-inexact-dual-ascent-x-x-prime-randomized}
  Consider the distributed AL algorithm with randomized Gauss-Seidel primal updates,
   where the expected number of inner iterations equals~$\tau$. Further, let Assumptions~\ref{assumption-bdd-hessian} and~\ref{assumption-network} hold.    Then, for all $k=0,1,\cdots$:
   \begin{align*}
  \mathbb E \left[\,\left\| x(k+1)-x^\prime(k+1)  \right\| \,\right]\leq e^{-\eta\,\tau} \, \mathbb E\left[\,\|x(k)-x^\prime(k+1) \|\,\right],
  \end{align*}
  where
  \begin{equation}
  \label{eqn-delta-prime-def-randomized}
  \eta:=N\,\left\{1-\left[\, 1-\frac{1}{N}\left(  1-\frac{\rho^2}{(\rho+h_{\mathrm{min}})^2}\right)  \,\right]^{1/2}\right\}.
  \end{equation}
  \end{lemma}
\begin{IEEEproof}
Fix some $k$, fix some $j=1,2,...$, and take $\omega \in \mathcal{A}_{k,j}$.
Thus, $\tau(k)=\tau(k;\omega)=j$ and there are $j$ inner iterations.
Fix some $s$, $s \in \{0,1,...,j-1\}$, and suppose that $\widehat{\imath}(k,s)=i$ (node $i$ is activated.)
We have that $x_i(k,s+1)$ satisfies the following:
\begin{equation*}
x_i(k,s+1) = \Phi_{i}^{-1} \left(  \sum_{j \in {O_i}}\,\rho\,W_{ij}\,x_j(k,s)-{{{\mu}}}_i(k)  \right).
\end{equation*}
On the other hand, we know that $x_i^\prime(k+1)$ satisfies:
\begin{equation*}
x_i^\prime(k+1) = \Phi_{i}^{-1} \left(  \sum_{j \in {O_i}}\,\rho\,W_{ij}\,x_j^\prime(k+1)-{{{\mu}}}_i(k)  \right).
\end{equation*}
Subtracting the above equalities, and using~\eqref{eqn-phi-inv-bounds-i},
letting
\begin{align*}
&R_{\Phi,i}(s):=R_{\Phi,i}(\rho\,\sum_{j \in {O_i}}\,W_{ij}\,x_j(k,s)-{{{\mu}}}_i(k)\,,\\
&
\,\rho\,\sum_{j \in {O_i}}\,W_{ij}\,x_j^\prime(k+1)-{{{\mu}}}_i(k)),
\end{align*}

and squaring the equality, we obtain:
{
\allowdisplaybreaks
\begin{eqnarray}
&\,& \left(x_i(k,s+1)-x_i^\prime(k+1)\right)^2 \nonumber\\
&=& \left(R_{\Phi,i}(s)\right)^2 \rho^2\,\left(\sum_{j \in {O_i}}\, W_{ij}\,(x_j(k,s)-x^\prime_j(k+1))\right)^2 \nonumber
\\
\label{eqn-ineq-convexity-randomized}
&\leq&
\left(\frac{\rho}{\rho+h_{\mathrm{min}}}\right)^2 \sum_{j \in {O_i}}\, W_{ij}\,(x_j(k,s)-x^\prime_j(k+1))^2\\
&=&
\label{eqn-ineq-convexity-randomized-2}
\delta^2\,\sum_{j=1}^N \,W_{ij}\,\,(x_j(k,s)-x^\prime_j(k+1))^2.
\end{eqnarray}}
Here,~\eqref{eqn-ineq-convexity-randomized}
further uses: 1) convexity of the quadratic function $u \mapsto u^2$; 2) the fact that
$\sum_{j \in {O_i}} W_{ij}=1$; and 3) the fact that the $W_{ij}$'s are nonnegative.
Also,~\eqref{eqn-ineq-convexity-randomized-2} introduces notation:
 $\delta:=\frac{\rho}{\rho+h_{\mathrm{min}}}$,
 and uses the fact that $W_{ij}=0$ if $\{i,j\} \notin E$ and $i \neq j$.
As node $i$ is selected, the remaining
quantities $x_j(k,s)$, $j \neq i$, remain
unchanged; i.e., $x_j(k,s+1)-x^\prime_j(k+1)=x_j(k,s)-x^\prime_j(k+1)$, $j \neq i$.
 Squaring the latter equalities, adding them up for all $j \neq i$,
 and finally adding them to~\eqref{eqn-ineq-convexity-randomized-2}, we obtain:
\begin{eqnarray}
&\,&\|x(k,s+1)-x^\prime(k+1)\|^2 \nonumber\\&\leq&
\|x(k,s)-x^\prime(k)\|^2 \nonumber\\
&+&
\delta^2\,\sum_{j=1}^N W_{ij}\,(x_j(k,s)-x_j^\prime(k+1))^2 \nonumber \\
&-& (x_i(k,s)-x_i^\prime(k+1))^2,
\label{eqn-adding-up-conditional-ineq}
\end{eqnarray}
for any $\omega \in \mathcal{A}_{k,j}$ such that $\widehat{\imath}(k,s)=i$.

We now compute conditional expectation of $\|x(k,s+1)-x^\prime(k+1)\|^2$,
conditioned on $\tau(k)=j$, $x(k)=x(k,0)$, ${\mu}(k)$,
and $x(k,1),...,x(k,s),$ $s \leq j-1.$ Conditioned on the latter,
each node $i$ updates equally likely, with conditional probability
 $1/N$, and therefore:
 {
\allowdisplaybreaks
 %
%
\begin{align}
\nonumber
& \mathbb E \left[\, \|x(k,s+1)-x^\prime(k+1)\|^2\, \,|\,
x(k),{\mu}(k),\tau(k)\right.
\\
\nonumber
&
\left.=j ,x(k,1),...,x(k,s)\,\right]
\\
&
\nonumber
 \leq  \|x(k,s)-x^\prime(k+1)\|^2 +\\
 \nonumber
&
\nonumber
\frac{1}{N}\,\delta^2\,\sum_{i=1}^N \sum_{j=1}^N W_{ij}\,(x_j(k,s)-x_j^\prime(k+1))^2  \\
&
\nonumber
- \frac{1}{N}\sum_{i=1}^N(x_i(k,s)-x_i^\prime(k+1))^2  \\
&
\nonumber
=
\|x(k,s)-x^\prime(k+1)\|^2 \\
&
+
\frac{1}{N}\,\delta^2\,\sum_{i=1}^N W_{ij}\,\sum_{j=1}^N \,(x_j(k,s)-x_j^\prime(k+1))^2  \nonumber \\
 &
 - \frac{1}{N}\|x(k,s)-x^\prime(k+1)\|^2 
\label{eqn-adding-up-conditional-ineq-expectation} \\
\nonumber
&
=\!\! \|x(k,s)-x^\prime(k+1)\|^2
\!+\!\frac{1}{N}\delta^2\|x(k,s)\!-\!x^\prime(k+1)\|^2  \nonumber \\
\label{eqn-adding-up-conditional-ineq-expectation-3}
&
- \frac{1}{N}\|x(k,s)-x^\prime(k+1)\|^2,
\:\forall \omega \in {\mathcal A}_{k,j}.
\end{align}
}
Here, inequality~\eqref{eqn-adding-up-conditional-ineq-expectation-3} uses
the fact that $\sum_{i=1}^N \,W_{ij}=1,$ $\forall j$. Rewriting~\eqref{eqn-adding-up-conditional-ineq-expectation-3},
we get:
\begin{align*}
&\mathbb E \left[ \left.\|x(k,s+1)-x^\prime(k+1)\|^2\right|\right. \\
&
\left.
\hspace{1cm}
\left|
\phantom{{x}^2}
\hspace{-.35cm}
x(k),{\mu}(k),\tau(k)=j,x(k,1),...,x(k,s-1)\right.\right]
\\
&\leq\!\! \left( 1 - \frac{1}{N}(1-\delta^2) \right)\!\!\left\| x(k,s)\!-\!x^\prime(k+1) \right\|^2\!,
\forall \omega \in {\mathcal A}_{k,j}.
\end{align*}
Denote by $\delta^\prime:=\left( 1 - \frac{1}{N}(1-\delta^2) \right)^{1/2}$.
Using the Jensen inequality for quadratic convex functions and conditional expectation:
 $\mathbb E [U^2\,|\,V] \geq \mathbb E^2[|U|\,\,|\,V]$, we obtain:
\begin{align*}
&
\mathbb E \left[ \|x(k,s+1)-x^\prime(k+1)\| | x(k),{\mu}(k),\tau(k)=j,\right.\\
&
\hspace{4cm}
\left.x(k,1),...,x(k,s-1) \right]
\\
&
\leq \delta^\prime \left\| x(k,s)-x^\prime(k+1) \right\|,\forall \omega \in \mathcal{A}_{k,j}.
\end{align*}
Integrating with respect to $x(k,1),...,x(k,s)$:
\begin{align*}
&\mathbb E \left[\|x(k,s+1)-x^\prime(k+1)\| | x(k),{\mu}(k),\tau(k)=j\right]
\\
&
\leq \delta^\prime \mathbb E \left[ \left\| x(k,s)-x^\prime(k+1) \right\| | x(k),{\mu}(k),\tau(k)=j \right],\\
&
\hspace{6cm}
\forall \omega \in \mathcal{A}_{k,j}.
\end{align*}
Applying the above inequality for $s=0,1,...,j-1$, and using $x(k,s=\tau(k)=j)=x(k+1)$:
\begin{align*}
&\mathbb E \left[\, \|x(k+1)-x^\prime(k+1)\| \,|\, x(k),{\mu}(k),\tau(k)=j\,\right]
\\
&
 \leq (\delta^\prime)^j \mathbb E \left[ \left\| x(k)-x^\prime(k+1) \right\| | x(k),{\mu}(k),\tau(k)=j \right],\\
 &
 \hspace{4.5cm}
 \forall \omega \in \mathcal{A}_{k,j},\forall j=0,1,...,
\end{align*}
and so:
\begin{align*}
&\mathbb E \left[\, \|x(k+1)-x^\prime(k+1)\| \,|\, x(k),{\mu}(k),\tau(k)\,\right]
\\
&
 \leq (\delta^\prime)^{\tau(k)} \mathbb E \left[ \left\| x(k)-x^\prime(k+1) \right\| | x(k),{\mu}(k),\tau(k) \right],\mathrm{a.s.}
\end{align*}
Integrating with respect to $x(k),{\mu}(k)$:
\begin{align*}
&\mathbb E \left[\|x(k+1)-x^\prime(k+1)\| | \tau(k)\,\right]
\\
& \leq (\delta^\prime)^{\tau(k)} \mathbb E \left[ \left\| x(k)-x^\prime(k+1) \right\| | \tau(k) \right] \\
&
=
  (\delta^\prime)^{\tau(k)} \mathbb E \left[ \left\| x(k)-x^\prime(k+1) \right\|  \right]  ,\mathrm{a.s.,}
\end{align*}
where we used independence of $\tau(k)$ and $x(k),{\mu}(k)$.
   Taking expectation,  we obtain:
\begin{align*}
&\mathbb E \left[\, \|x(k+1)-x^\prime(k+1)\| \right]
\\
&
\leq \mathbb E [\,(\delta^\prime)^{\tau(k)}\,] \,\, \mathbb E \left[ \left\| x(k)-x^\prime(k+1) \right\|  \right].
\end{align*}
%
Because $\tau(k)$ is distributed
according to the Poisson distribution with parameter $N\,\tau$, we have:
 $\mathbb E \left[ (\delta^\prime)^{\tau(k)} \right]=
\sum_{l=0}^{\infty} (\delta^\prime)^l \frac{e^{-N \tau} (N \tau)^l}{l\!}=e^{-(1-\delta^\prime)N\,\tau}$.
We get:
\begin{align}
\label{eqn-expectation-inequality}
&\mathbb E \left[\, \|x(k+1)-x^\prime(k+1)\| \,\right] \leq
\\
&
\nonumber
e^{-(1-\delta^\prime)N\,\tau}\,\mathbb E \left[ \left\| x(k)-x^\prime(k+1) \right\|\right].
\end{align}
Substituting the expression for $\eta$, we obtain the desired result.
\end{IEEEproof}
Consider Theorem~1. Note that it does not apply directly to
the randomized algorithm variants. However, it can be easily adapted to the randomized variants as well. Namely,
consider the following random inexact AL method. Use the same initialization as
for~\eqref{eqn-update-primal-NEW}--\eqref{eqn-update-dual-NEW}. Given $x(k)$, $\mu(k)$,
define (as before) $x^\prime(k+1):=x^\prime(\mu(k)):=\mathrm{arg\,min}_{x}L_a(x;\mu(k))$.
 The primal update is as follows: let $x(k+1)$ be a random variable that obeys
$\mathbb E [\|x(k+1)-x^\prime(k+1)\|]\leq \xi\,\mathbb E [\|x(k)-x^\prime(k+1)\|]$. (This replaces~\eqref{eqn-update-primal-NEW} in Theorem~1.)
The dual update is the same as in~\eqref{eqn-update-dual-NEW}.
 Then, it is straightforward to show that, under condition~(16), the following holds:
  $\mathbb E [\,\|x_i(k)-x^\star\|\,] $ $\leq r^{k}$
   $\sqrt{N} \max \left\{ {D_x},\frac{2{D_{\mu}}}{\sqrt{\lambda_2(\mathcal L)}h_{\mathrm{min}}} \right\},$
    where $r$ is in~(17). Now, applying Lemma~7, the last result holds for the randomized Gauss-Seidel variant, with
 $\xi=e^{-\eta\,\tau}$. It turns out that an analogous conclusion also holds for the randomized gradient variant, with $\eta$ relaced by
   $\eta^\prime$, defined in the following Lemma.

\begin{lemma}[Randomized gradient]
  \label{lemma-inexact-dual-ascent-x-x-prime-randomized-grad}
  Consider the distributed AL algorithm with randomized gradient primal updates, let
   the expected number of inner iterations equal~$\tau$, and ler the
   primal step-size $\beta \leq 1/(h_{\mathrm{max}}+\rho)$. Further, let Assumptions~\ref{assumption-bdd-hessian} and~\ref{assumption-network} hold.
   Then, for all $k=0,1,\cdots$:
   \begin{align*}
  \mathbb E \left[\left\| x(k+1)-x^\prime(k+1)  \right\| \right]
  \leq e^{-\eta^\prime\tau}  \mathbb E\left[\|x(k)-x^\prime(k+1) \|\right],
  \end{align*}
  where
  \begin{equation}
  \label{eqn-delta-prime-def-randomized-grad}
  \eta^\prime:=N\left\{1-\left[1-\frac{1}{N}\beta h_{\mathrm{min}}(1-\beta h_{\mathrm{min}})  \right]^{1/2}\right\}.
  \end{equation}
  \end{lemma}
The proof of Lemma~\ref{lemma-inexact-dual-ascent-x-x-prime-randomized-grad} is similar
to that of Lemma~\ref{lemma-inexact-dual-ascent-x-x-prime-randomized}.
 For the randomized algorithm and gradient updates,~\eqref{eqn-ineq-convexity-randomized}--\eqref{eqn-ineq-convexity-randomized-2}
  hold with $\frac{\rho^2}{(\rho+h_{\mathrm{min}})^2}$
   replaced by $(1-\beta\,h_{\mathrm{min}})^2.$

%

\section{Simulation example}
\label{section-simulation-examples}
We provide a simulation example with $l_2$-regularized logistic losses. The simulations corroborate a globally linear convergence for
 both the deterministic and randomized distributed AL methods, and show that it is usually advantageous to take a small number of inner iterations~$\tau$.

\textbf{Optimization problem}. We detail the simulation. We consider distributed learning via the $l_2$-regularized logistic loss; see, e.g.,~\cite{BoydADMoM} for further details.
Nodes minimize the logistic loss:
\[
\sum_{i=1}^N f_i(x)=\sum_{i=1}^N \left(\,\log\left(  1+e^{-b_{i} (a_{i}^\top x_1 + x_0)}\right) + \frac{\mathcal{P}\,\|x\|^2}{2\,N}\right),
\]
%
where $\mathcal{P}>0$ is the regularization parameter, $x=(x_1^\top,x_0)^\top \in {\mathbb R}^{15}$, $a_{i} \in {\mathbb R}^{14}$ is the node $i$'s feature vector, and
$b_{i} \in \{-1,+1\}$ is its class label.
The Hessian $\nabla^2 f_i(x) = \frac{\mathcal{P}}{N}\,I\,+
\frac{e^{-c_i^\top x}}{(1+e^{-c_i^\top x})^2} c_i c_i^\top$,
where $c_i=(b_i a_i^\top, b_i)^\top \in {\mathbb R}^{15}$.
We take node $i$'s constants $h_{\mathrm{min},i}$
  and $h_{\mathrm{max},i}$ as:
   $h_{\mathrm{min},i} = \frac{\mathcal{P}}{N}$
    and $h_{\mathrm{max},i} = \frac{\mathcal{P}}{N} + \frac{1}{4}\,\|c_i\,c_i^\top\|$. (Note that $\frac{e^{-c_i^\top y}}{(1+e^{-c_i^\top y})^2}\leq 1/4$ for all $y$.) Further, we let $h_{\mathrm{min}}=\min_{i=1,\cdots,N}h_{\mathrm{min},i}$
     and $h_{\mathrm{max}}=\max_{i=1,\cdots,N}h_{\mathrm{max},i}$.
     For the specific problem instance here, the condition number $\gamma=h_{\mathrm{max}}/h_{\mathrm{min}}=49.55.$

\textbf{Data}. The $a_{i}$'s are independent over $i$. Their entries and the entries of the ``true'' vector ${{x^\star}}=({x_1^\star}^\top, {x_0^\star})^\top$ are independent standard normal. The class labels are
$
b_{i}=\mathrm{sign}
\left( {{{x^\star}}_1}^\top a_{i}+{{x^\star}}_0+\epsilon_{i}\right),
$
where the $\epsilon_{i}$'s are independent zero mean, standard deviation~$0.001$, Gauss.

\textbf{Network}. The network is geometric, $10$~nodes
placed uniformly randomly on a unit square, connected by an edge (28 links) if their distance
less than a radius.

\textbf{Algorithm parameters, metrics, and implementation}. We set the weight matrix $W=\frac{1.1}{2}I+\frac{0.9}{2}W_{m}$,
where $W_m$ is the Metropolis weight matrix. (Note that $W \succ 0$.)
Further, $\alpha=\rho=h_{\mathrm{min}}$
with all algorithm variants, and $\beta=\frac{1}{\rho+h_{\mathrm{max}}}=\frac{1}{(\gamma+1)h_{\mathrm{min}}}$
with the methods that use the gradient primal updates.
For the deterministic variant and Jacobi updates, we set the number of inner iterations $\tau
= \left\lceil \frac{\log \left( \frac{3(1+\gamma)} {\lambda_2(\mathcal L)}\right)}
{\log(2)} \right\rceil$
;
with the deterministic gradient variant~$\tau=
 \left\lceil \frac{ \log \left( \frac{3(1+\gamma)} {\lambda_2(\mathcal L)}\right)}
{\log\left( \frac{\gamma+1}{\gamma} \right)} \right \rceil $;
with the randomized Gauss-Seidel variant~$\tau
=\left\lceil \frac{ \left| \log \left(  \frac{3(1+\gamma)}{\lambda_2(\mathcal L)} \right)  \right|   }{N\,\left( 1-(1-3/(4\,N))^{1/2}\right)  } \right\rceil
$;
and with the randomized gradient variant~$\tau
 = \left\lceil \frac{ \left| \log \left(  \frac{3(1+\gamma)}{\lambda_2(\mathcal L)}\right)  \right|   }{N\,\left( 1-
\left(1-\frac{\gamma}{N(1+\gamma)^2}\right)^{1/2}\right)  } \right \rceil
$. The above values of the algorithm parameters
$\alpha,\beta,\rho$, and $\tau$ satisfy conditions of Theorem~1 and Lemmas~5--8, and hence
they guarantee linear convergence rates.
We also simulate the methods with $\tau=1$ (although our theory does not guarantee linear convergence in such case.)
We initialize from zero the primal and dual variables with all methods.
We consider $\frac{1}{N}\sum_{i=1}^N \frac{f(x_i)-f^\star}{f(0)-f^\star}.$
We compare the methods in terms of: 1)~total number of transmissions (across all nodes),
and 2)~total computational time.
We implement the methods via a serial implementation -- one processor works
the jobs of all nodes. We count the CPU time for the overall jobs across all nodes.
With the methods that use the Gauss-Seidel and Jacobi updates in~\eqref{eqn-AL-algorithm-1},
we solve the local problems via the fast Nesterov gradient method for strongly convex functions.
At the inner iteration $s$ and outer iteration $k$, to solve~\eqref{eqn-AL-algorithm-1},
we initialize the Nesterov gradient method by $x_i(k,s)$.
We stop the algorithm after:
$
\left\lceil \left| \frac{ \log\left(\frac{2\epsilon }{(R^\prime)^2 L^\prime} \right)}{ \log(  1-\sqrt{\gamma^\prime})} \right| \right\rceil
$
 iterations, with\footnote{We implicitly assume that
the physical time allocated for each inner iteration $s$ suffices
to perform optimization~\eqref{eqn-AL-algorithm-1}.} $\epsilon=10^{-5}$. This guarantees
that the optimality gap upon termination is below~$\epsilon=10^{-5}$.
Here, $L^\prime$ is a Lipschitz constant for the cost
function in~\eqref{eqn-AL-algorithm-1} that (at node $i$) we take as $h_{\mathrm{max},i}+\rho+\frac{\mathcal R}{N}.$
 Further, $\gamma^\prime=L^\prime/\nu^\prime$ is the cost condition number, where $\nu^\prime = \frac{\mathcal R}{N}+\rho$ is the Hessian lower bound. The estimate of the distance
 to the solution is $R^\prime
  = \frac{1}{\rho+\mathcal R/N} \| \nabla f_i(x_i(k,s))  + (\mathcal R/N + \rho) x_i(k,s) + \left(\mu_i(k) - \rho \overline{x}_i(k,s)\right) \|.$
 All Figures are in semi-log scale.

In Figure~1~(top left), 
  we plot the relative error in the cost function for the deterministic variants
versus the number of communications, while in Figure~1 (top right), we depict
 the same quantity versus the CPU time
(This is the cumulative CPU time across all nodes.)
 We simulate the Jacobi method with both theoretical value of~$\tau$ and $\tau=1$, and the gradient method
 with both theoretical value of~$\tau$ and $\tau=1$.
 The Figures illustrate the linear convergence of the proposed methods. We report that the gradient
method with the theoretical value of~$\tau$ also shows a linear
convergence in the number of communications, but it converges slowly due to the large value of $\tau$.
 The Jacobi variant is better in terms of
communication cost but is worse in terms of computational cost.
\begin{figure}[thpb]
      \centering
       \includegraphics[height=2.2 in,width=2.9 in]{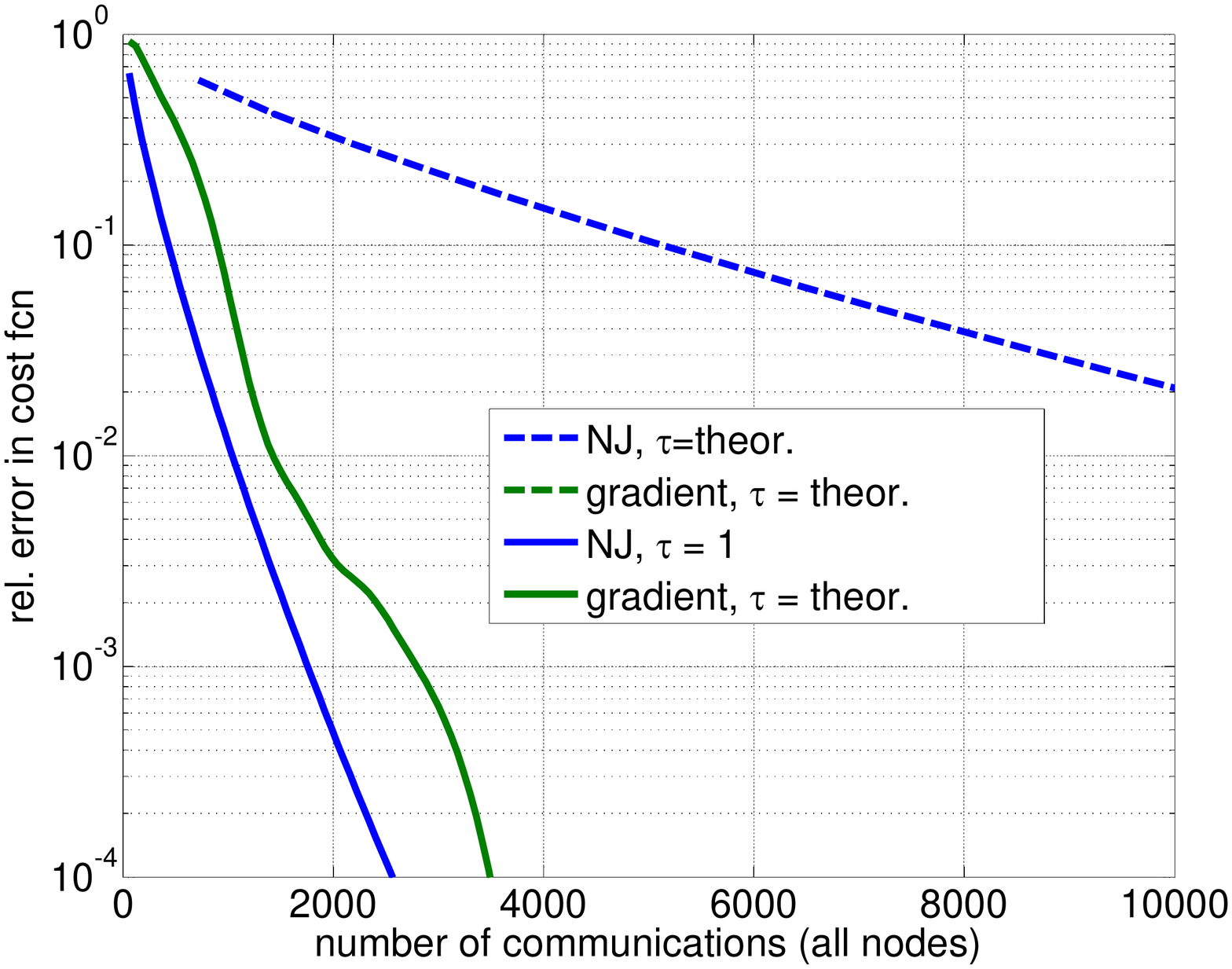}
       \includegraphics[height=2.2 in,width=2.9 in]{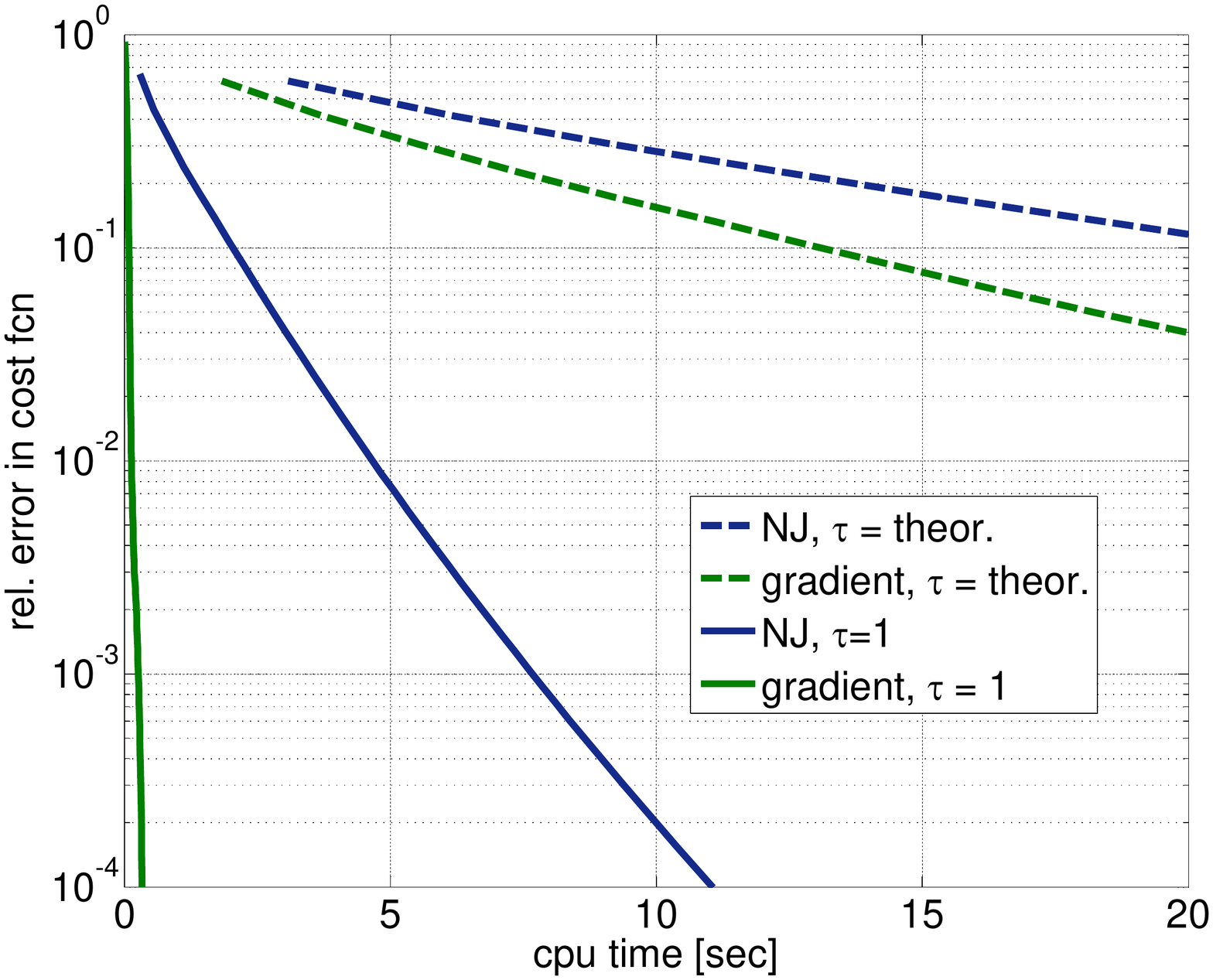}
       \includegraphics[height=2.2 in,width=2.9 in]{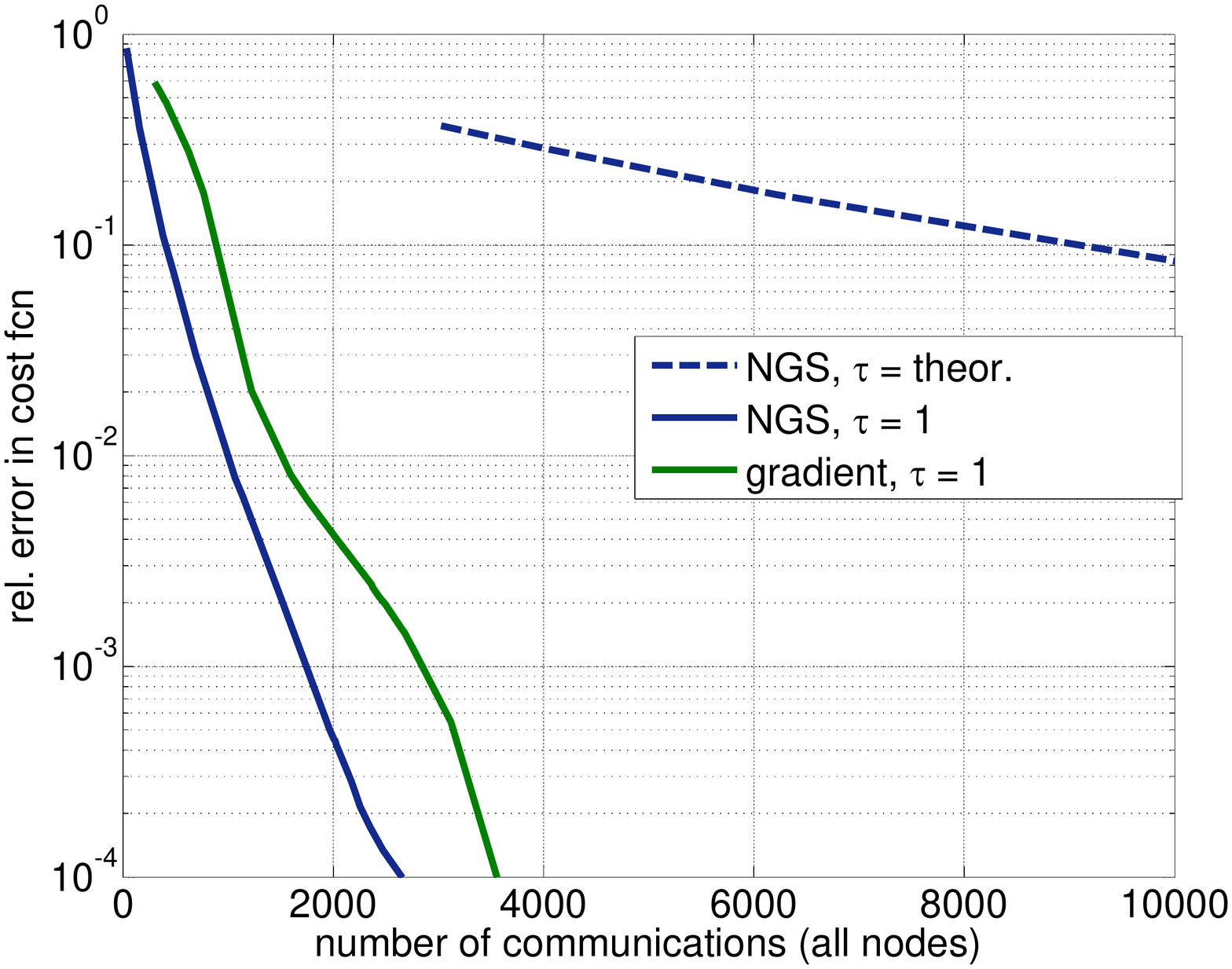}
       \includegraphics[height=2.2 in,width=2.9 in]{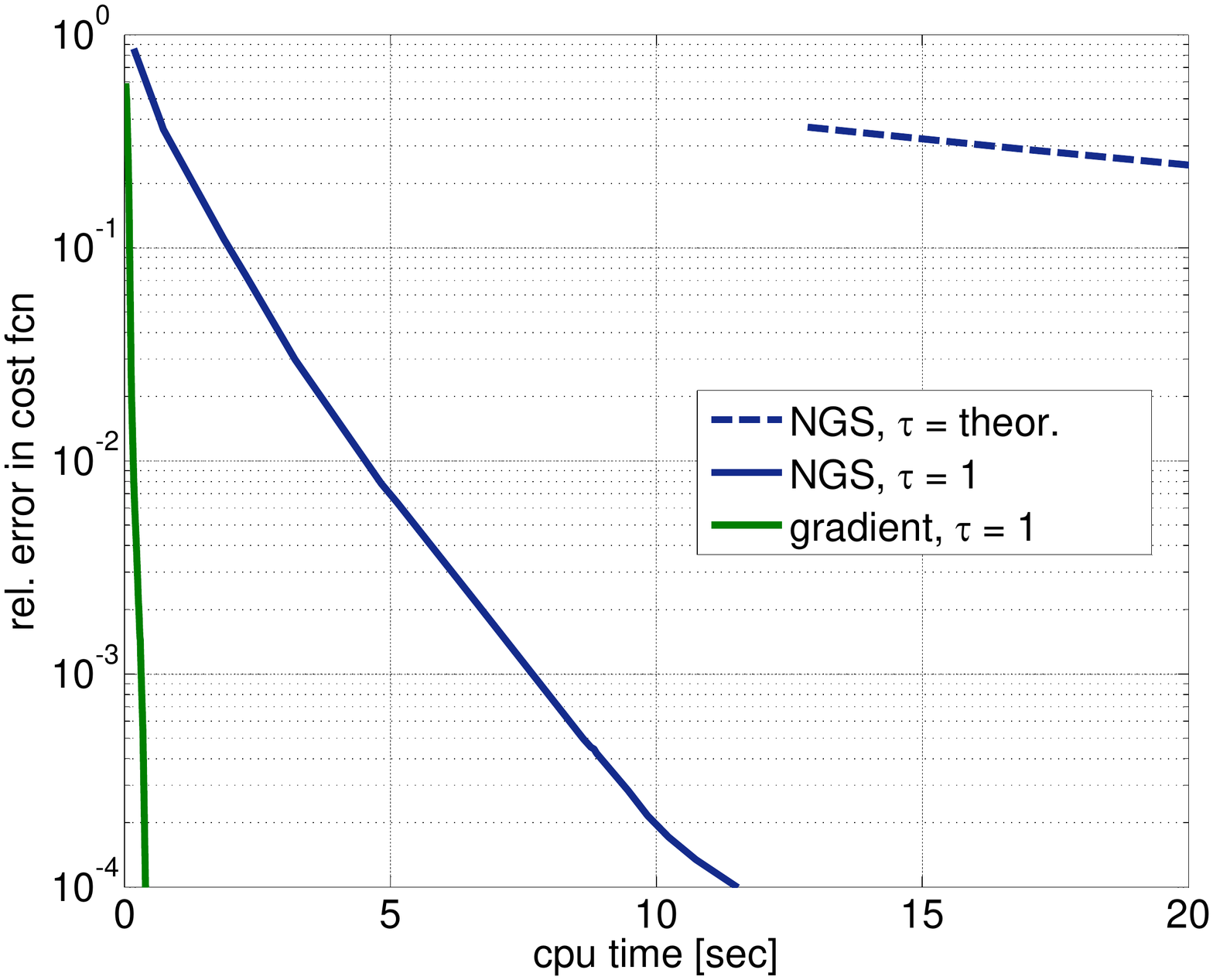}
       \caption{Deterministic (two left most) and randomized (two most right) AL methods: Average relative error in the cost function
       $\frac{1}{N}\sum_{i=1}^N \frac{f(x_i)-f^\star}{f(0)-f^\star}$. First and third plots: communication cost (total number
        of communications across all nodes). Second and fourth plots: computational cost (total CPU time across all nodes.) NJ--Jacobi; NGS--Gauss-Seidel.}
\end{figure}
Figures~1 (bottom left and right)
  present the same plots for the randomized Gauss-Seidel and gradient methods.
 The behavior is similar to the deterministic variants.
 The theoretical value for $\tau$ of the randomized gradient method is very large, and, consequently, the algorithm shows slow convergence for the latter choice of~$\tau$.

\vspace{-4mm}
\section{Conclusion}
\label{section-conclusion}
 We consider distributed optimization where $N$ nodes minimize the sum
 of their convex costs $f_i$'s by four
distributed augmented Lagrangian (AL) methods that differ in the primal variable updates: 1)
 deterministic AL with Jacobi updates; 2)
 deterministic AL with gradient descent; 3) randomized AL with nonlinear Gauss-Seidel;
 and 4) randomized AL with gradient descent updates. With twice continuously
 differentiable costs with bounded Hessian, we establish globally linear (geometric) convergence rates
 for all methods and give explicit dependence of the rates on the underlying network parameters.
  Simulation examples demonstrate linear convergence of our methods.

\vspace{-7mm}

\bibliographystyle{IEEEtran}
\bibliography{IEEEabrv,bibliography_new}

\end{document}